\documentstyle[12pt]{article}

\makeatletter
\def\section{\@startsection{section}{1}{\z@}{-3.5ex plus -1ex minus -.2ex}
{2.3ex plus .2ex}{\large\bf}}
\makeatother

\makeatletter
\@addtoreset{equation}{section}

\makeatother

\parskip=\medskipamount

\begin{document}

\vspace{8mm}

\begin{center}

{\Large \bf Perturbative expansion in gauge theories on compact manifolds} \\

\vspace{12mm}

{\large David H. Adams\footnote{Supported by FORBAIRT scientific research
program SC/94/218.}}

\vspace{4mm}

School of Mathematics, Trinity College, Dublin 2, Ireland. \\

\vspace{1ex}

email: dadams@maths.tcd.ie

\end{center}

\begin{abstract}

A geometric formal method for perturbatively expanding functional integrals  
arising in quantum gauge theories is described when
the spacetime is a compact riemannian manifold without boundary.
This involves a refined version of the Faddeev-Popov procedure using the 
covariant background field gauge-fixing condition with background gauge field
chosen to be a general critical point for the action functional (i.e. a 
classical solution). The refinement takes into account the
gauge-fixing ambiguities coming from gauge transformations which leave the 
critical point unchanged, resulting in the absence of infrared divergences 
when the critical point is isolated modulo gauge transformations.
The procedure can be carried out using only the 
subgroup of gauge transformations which are topologically trivial, possibly
avoiding the usual problems which arise due to gauge-fixing ambiguities.
For Chern-Simons gauge theory the method enables the partition function
to be perturbatively expanded for a number of simple spacetime manifolds
such as $S^3$ and lens spaces, and the expansions are shown to be formally
independent of the metric used in the gauge-fixing.

\end{abstract}

\newpage

\section{Introduction}

The functional integrals associated with vacuum expectation values 
in gauge theories have a geometric nature which
allows them to be formulated in the setting where the spacetime has
curved geometry and non-trivial topology, and where the gauge fields are 
associated with a topologically non-trivial principal fibre bundle.
There are a number of reasons why it is interesting to
consider these functional integrals in this general setting:
(i) The curved spacetime is equivalent to a background gravitational field
(when the geometry of the spacetime is pseudo-riemannian). 
(ii) Yang-Mills gauge theory on Euclidean ${\bf R}^4$ is equivalent 
(with regards to finite action gauge fields) 
to Yang-Mills gauge theory on the compact 
riemannian manifold $S^4$ because of the conformal invariance of the theory,
and the gauge fields on ${\bf R}^4$ satisfying topologically twisted
boundary conditions at $\infty$ are associated with non-trivial principal
fibre bundles over $S^4\,$
(see e.g. \cite{Sch(Top)+NashSen} for a discussion of this).
(iii) In a topological gauge theory, the Chern-Simons theory, 
these integrals lead
to a new quantum field theoretic way of obtaining topological invariants
of compact 3-dimensional manifolds, and of linked knots embedded in these
manifolds, as discussed in \cite{Sch(Baku)} and later (independently) 
explicitly demonstrated in \cite{W(Jones)}.
In the usual setting where the spacetime is flat the functional integrals are
usually evaluated as a perturbation series in a coupling parameter.
There is a well-developed formalism for carrying out this perturbative 
expansion via Feynman diagrams, described e.g. in \cite{ItZu}. 
This formalism extends to the curved spacetime setting as discussed in
\cite{Davies}; however this presumes a trivialisation of the principal
fibre bundle with which the gauge fields are associated, since when the
bundle is non-trivial there is no canonical decomposition of the action
functional into a quadratic ``kinetic'' term and higher order ``interaction''
term.

Our aim in this paper is to provide an extension of the formalism for
perturbatively expanding the functional integrals arising in quantum gauge 
theories to the setting where the spacetime is a general riemannian manifold
$M$ and the gauge fields are associated with an arbitrary principal fibre
bundle $P$ over $M$ (i.e. the gauge fields are the connection 1-forms
on $P\,$). For technical reasons (discussed in the conclusion) we take the
spacetime manifold to be compact and without boundary.
This problem has been studied previously 
by S.~Axelrod and I.~Singer in the context of Chern-Simons gauge theory
with spacetime a general compact 3-dimensional manifold 
\cite{AxSi1}\footnote{Perturbative expansion in Chern-Simons gauge theory on
${\bf R}^3$ has been extensively studied in the physics literature, see 
\cite{Guad} and the references therein.
A rigorous treatment of the perturbative definition
of knot invariants in ${\bf R}^3$ up to two loops was given by 
D.~Bar-Natan in \cite{Natan}.}.
They decomposed the action functional into kinetic- and interaction terms
by expanding it about a classical solution, i.e. a flat gauge field, and
extended the standard formalism (with BRS gauge fixing) to perturbatively
expand the Chern-Simons partition function, showing that the expansion
was ultraviolet finite (with a natural point-splitting regularisation) and
(essentially) independent of the metric used in the gauge fixing. (A result
on the topological nature of the expansion was later extended in \cite{AxSi2}).
However, the perturbative expansion obtained from their method is infrared
divergent unless a very restrictive condition is satisfied by the flat gauge
field about which the action is expanded: it must be isolated modulo gauge
transformations and irreducible. Because of
this their method is not applicable for a number of simple spacetime 
manifolds such as $S^3$ and the lens spaces since all the flat gauge fields
on these manifolds are reducible. One of the motivations for this 
paper is to provide a method for perturbative expansion which in the context
of Chern-Simons gauge theory extends the one given in \cite{AxSi1} to 
obtain an expansion of the partition function which is infrared finite as
well as ultraviolet finite for simple manifolds such as $S^3$ and the lens 
spaces. This would open up the possibility of explicitly evaluating the
terms in the expansions for these manifolds and comparing with the expressions
obtained from the non-perturbative prescription of \cite{W(Jones)}.
(This would provide a very interesting test of perturbation theory; such
tests have already been successfully carried out in the semiclassical
approximation as we discuss in the conclusion).

The formalism for perturbative expansion in this paper is presented in a 
general context which encompasses both Yang-Mills- and Chern-Simons gauge
theories. We consider functional integrals of the form
\begin{eqnarray}
I(\alpha;f,S)=\int_{\cal A}{\cal D}A\,f(A)e^{-\frac{1}{\alpha^2}S(A)}
\label{1.1}
\end{eqnarray}
where the formal integration is over a space ${\cal A}$ of gauge fields $A$
on a compact riemannian manifold $M$ without boundary,
$f(A)$ and $S(A)$ are gauge-invariant functionals on ${\cal A}$ and $\alpha
\in{\bf R}$ is a coupling parameter. (The functional integral (\ref{1.1})
arises in connection with the vacuum expectation values of a functional $f$
in a gauge theory with action functional $S$). We will describe a method for
carrying out formal perturbative expansions of (\ref{1.1}) in $\alpha$ via
a new geometric version of Feynman diagrams analogous to the momentum
space version of Feynman diagrams used in the usual flat spacetime setting.
In order to obtain a decomposition of the action functional $S(A)$ into 
a ``kinetic term'' and ``interaction term'' we expand about a general
critical point $A^c$ (i.e. a classical solution); this gives
\begin{eqnarray}
S(A^c+B)=S(A^c)+<B,D_{A^c}B>+S_{A^c}^I(B)
\label{1.2}
\end{eqnarray}
where $D_{A^c}$ is an operator which is self-adjoint w.r.t. the inner product
$<\cdot,\cdot>$ in the space of fields $B\,$, and $S_{A^c}^I(B)$ is a 
``polynomial'' in $B$ with each term of order ${\ge}3$ in $B$. 
The quadratic- and higher order terms in (\ref{1.2}) will play the roles of
``kinetic''- and ``interaction terms'' respectively.
When the spacetime manifold is compact without boundary the spectrum of
the operator $D_{A^c}$ is discrete (for the cases that we are interested in),
and the discrete variable labelling the spectrum will play an analogous
role to the momentum vector in the flat spacetime setting for constructing
the Feynman diagrams.

To rewrite (\ref{1.1}) in a form which can be perturbatively expanded we
develop a refined version of the Faddeev-Popov gauge-fixing 
procedure \cite{FP}. This uses the covariant background field
gauge-fixing condition with background gauge
field chosen to be the critical point $A^c$ for $S$ in (\ref{1.2}).
Our method for perturbative expansion is formal in the sense that the
problem of ultraviolet divergences is not addressed (although these
divergences do not arise in Chern-Simons gauge theory with point-splitting
regularisation, due to a result 
in \cite{AxSi1}). However, the problem of infrared 
divergences is considered in detail. The main feature of our method (besides
its geometric nature) is that infrared divergences do not arise when the
critical point $A^c$ is isolated modulo gauge transformations.
(We also briefly sketch how it may be possible to extend the method to the
case where $A^c$ is a completely general critical point, but our 
arguments for this are incomplete). This is a consequence of our refinement
of the Faddeev-Popov procedure, which takes into account the gauge-fixing
ambiguities coming from the isotropy subgroup of $A^c\,$, i.e. the gauge
transformations which leave $A^c$ unchanged. 
It is only when $A^c$ is reducible, i.e. when the isotropy subgroup of
$A^c$ is non-trivial, that our refinement leads to a different result than
the usual procedure described with the covariant background field condition
in \cite{Rouet}.
Also, we point out, as was first noted in \cite{Rouet}, that the
gauge-fixing procedure can be carried out using only the subgroup of
gauge transformations which are topologically trivial. This avoids the
usual problems which arise due to gauge-fixing ambiguities, provided that all
ambiguities which do not come from the isotropy subgroup of $A^c$ come from
topologically non-trivial gauge transformations (which we will assume to be
the case).

In the context of Chern-Simons gauge theory our method extends the one of
Axelrod and Singer in \cite{AxSi1} to allow for reducible flat 
gauge fields $A^c\,$, providing an ultraviolet- and
infrared-finite method for perturbatively expanding the partition function
for a number of simple spacetime 3-manifolds such as $S^3$ and the lens spaces.
(We discuss this in more detail in \S5). We show that the 
perturbative expansion of the partition function obtained from our method
(with $A^c$ isolated modulo gauge transformations) is formally 
metric-independent. This extends a result in \cite{AxSi1}.

The contribution to
expectation values from field fluctuations about instantons in Yang-Mills
gauge theories with compact riemannian spacetime
was studied in \cite{Sch(Inst)}. In connection with
this a formula was derived for the weak coupling ($\alpha{\to}0$) limit of
(\ref{1.1}) in \cite[App. II]{Sch(Inst)}. (This formula was also used
in \cite{Sch(degen)} to obtain an expression for the semiclassical
approximation for the partition function of a gauge theory).
We find that the lowest order term in our perturbative expansion of (\ref{1.1})
reproduces this formula. This is reassuring, since the formula in 
\cite[App. II]{Sch(Inst)} was derived without gauge-fixing, whereas
our method does use gauge-fixing.

This paper is organised as follows.
In \S2 we explain the basic ideas behind the perturbative expansion of the
functional integral (\ref{1.1}), including the geometric version of the
Feynman diagrams. The precise relationship between infrared divergences,
non-existence of the propagator and gauge invariance is determined.
This shows precisely what it is that a gauge-fixing procedure needs to do
to ensure a well-defined propagator and avoid infrared divergences.
In \S3 we describe the gauge-theoretic setup to be used in the rest of the
paper, fixing notations and stating a few basic formulae that we will be
using. 
In \S4 we rewrite the functional integral (\ref{1.1}) using a refined 
version of the Faddeev-Popov gauge-fixing procedure to obtain an expression
which can be perturbatively expanded without infrared divergences by the
method described in \S2 (at least when $A^c$ is isolated modulo gauge
transformations).
In \S5 we specialise to Chern-Simons gauge theory. 
We show that our approach to perturbative expansion extends the
approach of \cite{AxSi1} to the case where $A^c$ is isolated modulo gauge 
transformations and show the formal 
metric-independence of the perturbative expansion of the partition function
in this case.
In \S6 we make some concluding remarks.
Most of what we do in \S2--\S4 is formal. It seems possible that parts of \S4 
can be made rigorous; the results in \cite{Viallet1} may be of use for this.

The method described in this paper was
discussed previously by
the author in the context of Chern-Simons gauge theory on $S^3$
in \cite{talk}. 
Features of the method in the general case,
and their connection with \cite{AxSi1} were
later pointed out in \cite{AdSe(hep-th)}. 
In a recent overview paper \cite{Ax(hep-th)} S.~Axelrod has announced that
he has extended his previous work with I.~Singer \cite{AxSi1}, \cite{AxSi2}
on Chern-Simons gauge theory
to the very general case where $A^c$ (in (\ref{1.2})) is only required
to belong to a smooth component of the moduli space of flat gauge fields. 
The details of the method and arguments used for this have
yet to appear (as far as we are aware), 
and we do not know to what extent they coincide with ours.

{\bf Note added}. When the background gauge field is irreducible
it was shown in \cite{Viallet(PLB)} that
there is a very interesting and deep relationship between the Faddeev-Popov
determinant and the natural metric on the orbit space of the gauge fields
(see also \cite{Viallet2}). We expect that this relationship will continue
to hold for reducible background gauge fields, 
with the Faddeev-Popov determinant replaced by our modified expression 
(the inverse of (\ref{4.16}) below), although we have yet to verify this.

\section{A general method for perturbative expansion}

In this section we describe a formal method for perturbatively expanding
the functional integral (\ref{1.1}) in a general setting where ${\cal A}$
is an arbitrary (infinite-dimensional) affine space modelled on a vectorspace
$\Gamma$ with inner product $<\cdot,\cdot>$. 
The functional $f$ may be complex-valued while $S$ may be real-valued or
purely imaginary-valued. (Unless stated otherwise we take $S$ to be 
real-valued in the following; the modifications required when $S$ is replaced
by $iS$ will be clear). The functionals 
are required to satisfy the following basic condition. 
(Examples of functionals $S$ satisfying the condition are the action 
functionals for Yang-Mills- and Chern-Simons gauge theories, 
given by (\ref{3.3a}) and (\ref{3.3b}) below; an example of functional 
$f$ satisfying the condition is the
Wilson loop functional given by (\ref{3.3c}) below).
For each critical point $A^c$ for $S$ $f(A^c+B)$ and $S(A^c+B)$ are 
``polynomials'' in $B\in\Gamma$. More precisely, the functionals can be 
expanded as
\begin{eqnarray}
S(A^c+B)=\sum_{k=0}^sS_{A^c}^{(k)}(B)\ \ \ ,\ \ \ \ f(A^c+B)=
\sum_{k=0}^{\infty}f_{A^c}^{(k)}(B)
\label{22.1}
\end{eqnarray}
with $2{\le}s<\infty\,$, $\ S_{A^c}^{(k)}(B)=S_{A^c}^{(k)}(B,
\dots,B)\;$ and $\ f_{A^c}^{(k)}(B)=f_{A^c}^{(k)}(B,\dots,B)\,$
where $S_{A^c}^{(k)}(B_1,\dots,B_k)$ and $f_{A^c}^{(k)}(B_1,\dots,B_k)$
are multilinear functionals of $B_1,\dots,B_k\in\Gamma$.
Note that $S_{A^c}^{(0)}(B)=S(A^c)\;$, $\,f_{A^c}^{(0)}(B)=f(A^c)\,$
and $S_{A^c}^{(1)}=0$ since $A^c$ is a critical point for $S$. 
Since $S_{A^c}^{(2)}(B)$ is a quadratic functional we can write
\begin{eqnarray}
S_{A^c}^{(2)}(B)=<B,D_{A^c}B>
\label{22.2}
\end{eqnarray}
where $D_{A^c}$ is a uniquely determined selfadjoint operator on $\Gamma$.
(If real-valued $S$ is replaced by $iS$ then we replace $D_{A^c}$
by $iD_{A^c}$ in (\ref{22.2})).
This leads to the expression (\ref{1.2}):
\begin{eqnarray}
S(A^c+B)=S(A^c)+<B\,,\,D_{A^c}B>_1+S_{A^c}^I(B)
\label{22.2.5}
\end{eqnarray}
where $S_{A^c}^I(B)=\sum_{k{\ge}3}S_{A^c}^{(k)}(B)$. 
We now choose a specific critical point $A^c$ for $S$ and change variables in
the integration in (\ref{1.1}) from $A\in{\cal A}$ to $B=A-A^c\in\Gamma$
to obtain the following expression for the functional integral:
\begin{eqnarray}
I(\alpha;f,S)&=&\int_{\Gamma}{\cal D\/}Bf(A^c+B)e^{-\frac{1}{\alpha^2}
S(A^c+B)} \nonumber \\
&=&e^{-\frac{1}{\alpha^2}S(A^c)}\int_{\Gamma}{\cal D\/}Bf(A^c+B)
e^{-\frac{1}{\alpha^2}(<B,D_{A^c}B>+S_{A^c}^I(B))}
\label{22.3}
\end{eqnarray}
In the perturbative expansion of (\ref{22.3}) the 
quadratic term $<B\,,\,D_{A^c}B>_1$ and higher order term $S_{A^c}^I(B)$
in the exponential
will play the roles of ``kinetic term'' and ``interaction term'' 
respectively. We change variables from $B$ to $B'=\frac{1}{\alpha}B$ in
(\ref{22.3}) to obtain
\begin{eqnarray}
I(\alpha;f,S)
=e^{-\frac{1}{\alpha^2}S(A^c)}\int_{\Gamma}{\cal D\/}({\alpha}B')
f(A^c+{\alpha}B')e^{-<B',D_{A^c}B'>-\frac{1}{\alpha^2}S_{A^c}^I({\alpha}B')}
\label{22.4}
\end{eqnarray}
We choose an orthonormal basis $\{B_j\}_{j=0,1,2,\dots}$ for $\Gamma$ and
set $b_j=<B',B_j>\,$, then $B'=\sum_jb_jB_j$ and from (\ref{22.1}) we
get the expansions
\begin{eqnarray}
f(A^c+\alpha\tilde{B}')&=&\sum_{k=0}^{\infty}\alpha^k\,\sum_{j_1,\dots,j_k}
f_{A^c}^{j_1{\cdots}j_k}b_{j_1}{\cdots}b_{j_k} 
\label{22.5} \\
\frac{1}{\alpha^2}S_{A^c}^I(\alpha\tilde{B}')
&=&\sum_{k=1}^{s-2}\alpha^k\sum_{j_1,\dots,j_{k+2}}
S_{A^c}^{j_1{\cdots}j_{k+2}}b_{j_1}{\cdots}b_{j_{k+2}} 
\label{22.6} 
\end{eqnarray}
where $f_{A^c}^{j_1{\cdots}j_k}:=f_{A^c}^{(k)}(B_{j_1},\dots,B_{j_k})$ and
$S_{A^c}^{j_1{\cdots}j_k}:=S_{A^c}^{(k)}(B_{j_1},\dots,B_{j_k})$.
Substituting (\ref{22.5}) in (\ref{22.4}) leads to
\begin{eqnarray}
I(\alpha;f,S)=e^{-\frac{1}{\alpha^2}S(A^c)} 
\sum_{N=0}^{\infty}\alpha^N\sum_{j_1,\dots,j_N}f_{A^c}^{j_1{\cdots}j_N}
G_{A^c}^{(N)}(\alpha;j_1,\dots,j_N) 
\label{22.7}
\end{eqnarray}
where
\begin{eqnarray}
G_{A^c}^{(N)}(\alpha;j_1,\dots,j_N)&=&\int_{\Gamma}{\cal D\/}({\alpha}B')
<B',B_{j_1}>\cdots<B',B_{j_k}>e^{-<B',D_{A^c}B'>-\frac{1}{\alpha^2}S_{A^c}^I
({\alpha}B')} \nonumber \\
& &\label{22.8}
\end{eqnarray}
The functions $G_{A^c}^{(N)}(\alpha;j_1,\dots,j_N)$ will play an analogous
role to the Greens functions for field theories on flat spacetime. 
To perturbatively expand (\ref{22.7}) we must perturbatively expand the
Greens functions (\ref{22.8}). To do this we introduce a variable 
$J\in\Gamma$ (the ``source'' variable for $B'$), set $J_j=<J,B_j>$ and
rewrite (\ref{22.8}) via functional derivatives:
\begin{eqnarray}
\lefteqn{G_{A^c}^{(N)}(\alpha;j_1,\dots,j_N)} \nonumber \\
&=&\frac{\partial^N}{{\partial}J_{j_1}\cdots{\partial}J_{j_k}}
\exp\Bigl(-\frac{1}{\alpha^2}S_{A^c}^I(\alpha\frac{\partial}{{\partial}J})
\Bigr)\int_{\Gamma}{\cal D\/}({\alpha}B')e^{-<B',D_{A^c}B'>+<B',J>}
\ \biggl|_{J=0} \nonumber \\
& &\label{22.9}
\end{eqnarray}
The r.h.s. of this expression is to be understood as follows. Writing 
$<B',J>=\sum_j<B',B_j>J_j$ we consider the integral as an infinite polynomial
in $\{J_j\}_{j=0,1,2,\dots}$. The functional derivative $\frac{1}{\alpha^2}
S_{A^c}^I(\alpha\frac{\partial}{{\partial}J})$ is then the partial derivative 
operator obtained by replacing the $b_j$'s in (\ref{22.6}) by $\frac{\partial}
{{\partial}J_j}$'s.
We change variables in the integral in (\ref{22.9}) from $B'$ back to the old
variable $B={\alpha}B'$ and evaluate the integral using the generalisation
of the formula
\begin{eqnarray}
\int_{-\infty}^{\infty}e^{-{\lambda}x^2+ax}dx=\Bigl(\frac{\lambda}{\pi}
\bigr)^{-1/2}e^{\frac{a^2}{4\lambda}}
\label{22.9.5}
\end{eqnarray}
to obtain
\begin{eqnarray}
\int_{\Gamma}{\cal D\/}({\alpha}B')e^{-<B',D_{A^c}B'>+<B',J>}
=\det\Bigl(\frac{1}{\pi\alpha^2}D_{A^c}\Bigr)^{-1/2}e^{\frac{1}{4}<J,
(D_{A^c})^{-1}J>}
\label{22.10}
\end{eqnarray}
(Of course, $D_{A^c}$ will have zero-modes in general so the r.h.s. of 
(\ref{22.10}) is ill-defined. In the case of gauge theories this problem
is circumvented using a gauge-fixing procedure as we will see in \S4).
Substituting (\ref{22.10}) in (\ref{22.9}) enables $G_{A^c}^{(N)}
(\alpha;j_1,\dots,j_N)$ to be perturbatively expanded via Feynman diagrams,
as we now discuss. In order to simplify the expressions we choose the o.n.b.
$\{B_j\}_{j=0,1,2,\dots}$ to consist of eigenvectors for $D_{A^c}$ 
such that\footnote{We are assuming that $D_{A^c}$ has discrete spectrum;
this is the case for Yang-Mills- and Chern-Simons gauge theories on compact
riemannian manifolds as we will see in \S4.}
\begin{eqnarray}
D_{A^c}B_j&=&\lambda(j)B_j \nonumber \\
0{\le}|\lambda(0)|\le\cdots|\lambda(j)|&{\le}&|\lambda(j+1)|\le\cdots
\to\infty\ \ \ \mbox{for}\ \ \ j\to\infty
\label{22.11}
\end{eqnarray}
Using $<J,(D_{A^c})^{-1}J>=\sum_j\frac{1}{\lambda(j)}J_j^2$ we write 
(\ref{22.9}) as 
\begin{eqnarray}
\lefteqn{G_{A^c}^{(N)}(\alpha;j_1,\dots,j_N)} \nonumber \\
&=&\det\Bigl(\frac{1}{\pi\alpha^2}D_{A^c}\Bigr)^{-1/2}\frac{\partial^N}
{{\partial}J_{j_1}\cdots{\partial}J_{j_N}} \nonumber \\
& &\times\;\exp\Bigl(-\sum_{k{\ge}1}\alpha^k\Bigl(\,\sum_{i_1,\dots,i_{k+2}}
S_{A^c}^{i_1{\cdots}i_{k+2}}\frac{\partial^{k+2}}{{\partial}J_{i_1}\cdots
{\partial}J_{i_{k+2}}}\Bigr)\Bigr)\exp\Bigl(\,\sum_{j=0}^{\infty}
\frac{1}{4\lambda(j)}J_j^2\Bigr)\ \biggl|_{J=0} \nonumber \\
& &\label{22.12}
\end{eqnarray}
From this we see that the Greens functions can be perturbatively expanded as
\begin{eqnarray}
G_{A^c}^{(N)}(\alpha;j_1,\dots,j_N)=\det\Bigl(\frac{1}{\pi\alpha^2}D_{A^c}
\Bigr)^{-1/2}\sum_{k{\ge}0}\alpha^kG_{A^c}^{(k,N)}(j_1,\dots,j_N)
\label{22.13}
\end{eqnarray}
where each term $\alpha^kG_{A^c}^{(k,N)}(j_1,\dots,j_N)$ is obtained by a
Feynman diagram technique. The building blocks of the diagrams, and the
factors which each of these contribute, are as follows:

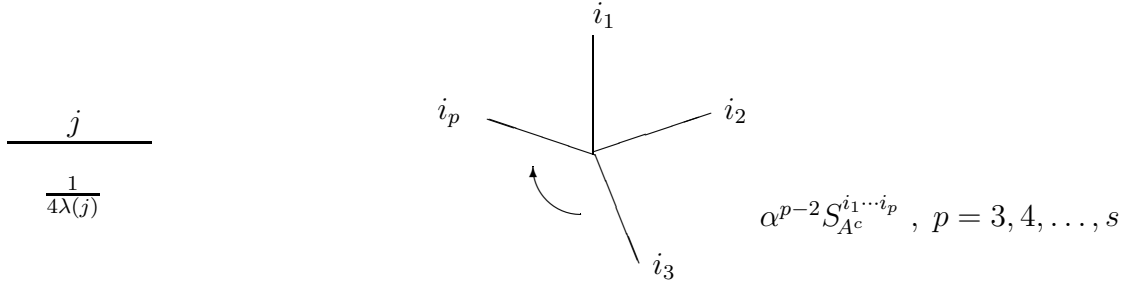
\setlength{\unitlength}{0.01250000in}%
\begingroup\makeatletter\ifx\SetFigFont\undefined%
\gdef\SetFigFont#1#2#3#4#5{%
  \reset@font\fontsize{#1}{#2pt}%
  \fontfamily{#3}\fontseries{#4}\fontshape{#5}%
  \selectfont}%
\fi\endgroup%
\begin{picture}(317,114)(73,719)
\thinlines
\put(295,765){\vector( 0, 1){0}}
\put(315,765){\oval( 40, 40)[bl]}
\put(320,770){\line( 0, 1){ 50}}
\put(321,770){\line(-3, 1){ 45}}
\put(320,771){\line( 3, 1){ 49.500}}
\put(321,770){\line( 2,-5){ 18.276}}
\put( 75,775){\line( 1, 0){ 60}}
\put(375,785){\makebox(0,0)[lb]{\smash{\SetFigFont{12}{14.4}{\rmdefault}{\mddefault}{\updefault}$i_2$
}}}
\put(320,825){\makebox(0,0)[lb]{\smash{\SetFigFont{12}{14.4}{\rmdefault}{\mddefault}{\updefault}$i_1$
}}}
\put(345,720){\makebox(0,0)[lb]{\smash{\SetFigFont{12}{14.4}{\rmdefault}{\mddefault}{\updefault}$i_3$
}}}
\put(255,785){\makebox(0,0)[lb]{\smash{\SetFigFont{12}{14.4}{\rmdefault}{\mddefault}{\updefault}$i_p$
}}}
\put( 90,750){\makebox(0,0)[lb]{\smash{\SetFigFont{12}{14.4}{\rmdefault}{\mddefault}{\updefault}$\frac{1}{4\lambda(j)}$
}}}
\put(100,780){\makebox(0,0)[lb]{\smash{\SetFigFont{12}{14.4}{\rmdefault}{\mddefault}{\updefault}$j$
}}}
\put(390,740){\makebox(0,0)[lb]{\smash{\SetFigFont{12}{14.4}{\rmdefault}{\mddefault}{\updefault}$\alpha^{p-2}S_{\!A^c}^{i_1{\cdots}i_p}\ ,\;p=3,4,\dots,s$
}}}
\end{picture}

\noindent Each diagram for 
$\alpha^kG_{A^c}^{(N,k)}(j_1,\dots,j_N)$ has $N$ external
lines labelled by $j_1,\dots,j_N$. The term associated with each diagram
(a function of $j_1,\dots,j_N$) is obtained by taking the product of all 
the factors associated with the lines and vertices of the diagram and summing
over all the values of the indices of the internal lines, and then dividing
by the symmetry factor of the diagrams (as described e.g. in 
\cite[\S6-1-1]{ItZu}). Then $\alpha^kG_{A^c}^{(N,k)}(j_1,\dots,j_N)$ is the
sum of all topologically distinct diagrams which are proportional to
$\alpha^k$.

These Feynman diagrams are analogous to the momentum space diagrams for
Greens' functions for field theories on flat spacetime ${\bf R}^4\,$:
The discrete index $j\in\{0,1,2,\dots\}$ is analogous to the momentum vector
$p\in{\bf R}^4\,$, and the factor $\frac{1}{4\lambda(j)}$ associated with a
line labelled by $j$
in the diagrams is analogous to the momentum space propagator.
The term in $\alpha^kG_{A^c}^{(N,k)}(j_1,\dots,j_N)$ corresponding to 
a given diagram with $q$ internal lines has the form
\begin{eqnarray}
\frac{1}{4^N\lambda(j_1)\cdots\lambda(j_N)}{\sum}c_{i_1{\cdots}i_q} 
\frac{1}{4^q\lambda(i_1)\cdots\lambda(i_q)}
\label{22.14}
\end{eqnarray}
(with summation over repeated indices) where $c_{i_1{\cdots}i_q}$ is the 
product of the vertex factors of the diagram together with 
the inverse of the
symmetry factor of the diagram. For the perturbative expansion to be 
meaningful the terms (\ref{22.14}) must be finite. There are two reasons
why (\ref{22.14}) may diverge. First, if $D_{A^c}$ has zero-modes then
$\lambda(j)$ is zero for sufficiently small $j$ (cf. (\ref{22.11})),
leading to divergence of (\ref{22.14}). We call divergences of this type
infrared divergences. Secondly, (\ref{22.14}) diverges if the summand
in (\ref{22.14}) does not converge quickly enough to zero when 
$i_1,\dots,i_q$ become large, or equivalently, if the divergence
$\lambda(j)\to\infty$ for $j\to\infty$ is not sufficiently rapid.
We call divergences of this type ultraviolet divergences.
We will study the problem of infrared divergences below, and show in
\S4 how they can be avoided in the case of gauge theories using a 
gauge-fixing procedure. To deal with the problem of ultraviolet divergences
methods of regularisation and renormalisation need to be developed.
We will not address this problem in this paper, but make two remarks 
which may be relevant in this context:

\noindent (i) There are general theorems which set lower bounds on the rate
of divergence $\lambda(j)\to\infty\,$, for $j\to\infty\,$, in many cases
of interest,
see e.g. \cite[\S1.5]{Gilkey}. These may be useful for establishing
general convergence criteria for the diagrams.

\noindent (ii) The Feynman diagrams in our approach do not have one of the 
significant features of the diagrams for field theories on flat spacetime,
namely there is no general analogue of momentum conservation at the 
vertices of the diagrams. Conservation of momentum at the vertices of 
Feynman diagrams for field theories on flat spacetime is intimately related
to the translation invariance of the kinetic term in the action functional
of the theory. This suggests that for 
field theories on compact curved spacetime for which the kinetic term 
in the action is symmetrical (i.e. invariant under a group of isometries
of the spacetime manifold)
there may be simplifying conditions analogous to momentum 
conservation at the vertices of the Feynman diagrams. An example of this is
when the spacetime is a compact group manifold and the 
kinetic term in the action of a field theory is
invariant under the action of the group on itself: In this case there are 
simplifying conditions analogous to (but weaker than) momentum conservation
at the vertices of the diagrams; these arise due to the orthogonality
relations between the characters of the irreducible representations of
the group \cite{unpublished}.

Substituting (\ref{22.13}) in (\ref{22.7}) we finally obtain the perturbative
expansion of the functional integral (\ref{1.1}):
\begin{eqnarray}
\lefteqn{I(\alpha;f,S)} \nonumber \\
&=&\det\Bigl(\frac{1}{\pi\alpha^2}D_{A^c}\Bigr)^{-1/2}e^{-\frac{1}{\alpha^2}
S(A^c)}\Big{\lbrack}f(A^c)+\sum_{k=1}^{\infty}\alpha^k\Bigl(\,\sum_{N=1}^k
\sum_{j_1,\dots,j_N}f_{A^c}^{j_1{\cdots}j_N}G_{A^c}^{(N,k-N)}(j_1,\dots,j_N)
\Bigr)\Big\rbrack \nonumber \\
& &\label{22.15}
\end{eqnarray}

We have seen that the presence of infrared divergences in the preceding
perturbative expansion correspond to the zero-modes of $D_{A^c}\,$, i.e.
the nullspace $\ker(D_{A^c})$ of $D_{A^c}$. We show below that 
$\ker(D_{A^c})$ is related to the critical point $A^c$ of $S$ as follows:
Let ${\cal C}$ denote the set of critical points for $S\,$, then
\begin{eqnarray}
T_{A^c}{\cal C}\subseteq\ker(D_{A^c})\ \ \ \mbox{and}\ \ \ T_{A^c}{\cal C}=
\ker(D_{A^c})\ \ \ \mbox{in the generic case.}
\label{22.16}
\end{eqnarray}
Here $T_{A^c}{\cal C}$ is the set of tangents to ${\cal C}$ at $A^c\,$, i.e.
\begin{eqnarray*}
T_{A^c}{\cal C}=\Big\{\,\frac{d}{dt}\biggl|_{t=0}A^c(t)\ \biggl|\ A^c(t)
\ \mbox{smooth curve in}\ {\cal C}\subset{\cal A}\ \mbox{with}\ A^c(0)=A^c
\Big\}\,.
\end{eqnarray*}
From (\ref{22.16}) we find the precise reason why infrared divergences
are unavoidable in gauge theories without gauge-fixing: If ${\cal A}$ is
a space of gauge fields and $S(A)$ is gauge invariant then ${\cal C}$
is gauge invariant, and in particular the orbit ${\cal G}{\cdot}A^c$
of the group ${\cal G}$ of gauge transformations through $A^c$ is contained
in ${\cal C}\,$, so the tangentspace $T_{A^c}({\cal G}{\cdot}A^c)$ to the 
orbit at $A^c$ is contained in $T_{A^c}{\cal C}$ and it follows from
(\ref{22.16}) that 
\begin{eqnarray}
T_{A^c}({\cal G}{\cdot}A^c)\subseteq\ker(D_{A^c})\,.
\label{22.17}
\end{eqnarray}
This shows that $\ker(D_{A^c})$ is necessarily non-vanishing for gauge 
theories. (In gauge theories the action of ${\cal G}$ on ${\cal A}$
does not have any fixed points, so $T_A({\cal G}{\cdot}A){\ne}0$ for
all $A\in{\cal A}$.) When $D_{A^c}$ is a positive operator, e.g. for
Yang-Mills gauge theories, the result (\ref{22.17}) can be obtained 
in a simple, direct way by a standard argument, see e.g. \cite{Rouet}.
This argument does not hold in general though, since it assumes that
$<B,D_{A^c}B>=0\;\Rightarrow\;D_{A^c}B=0\,$, which is only true if $D_{A^c}$
is positive. Our argument for (\ref{22.16}) and (\ref{22.17}) does not
require this assumption.

We show (\ref{22.16}) as follows\footnote{Our argument goes along similar
lines to an argument used in determining the dimensions of instanton
modulispaces in Yang-Mills gauge theories, see e.g. 
\cite[Part IV]{BoossBleecker}.}. Given $A^c\in{\cal C}$ any critical point for
$S$ can be written as $A^c+B$ and is characterised by
\begin{eqnarray}
\frac{d}{dt}\biggl|_{t=0}S(A^c+B+tC)=0\ \ \ \mbox{for all }\ C\in\Gamma
\label{22.18}
\end{eqnarray}
From (\ref{22.1}) we see that the functional $C\to\frac{d}{dt}\Bigl|_{t=0}
S(A^c+B+tC)$ is linear, and can therefore be written as
\begin{eqnarray}
\frac{d}{dt}\biggl|_{t=0}S(A^c+B+tC)=<C,R_{A^c}(B)>_1
\label{22.19}
\end{eqnarray}
It follows from (\ref{22.1}) and (\ref{22.2}) that
\begin{eqnarray}
R_{A^c}(B)=2D_{A^c}(B)+\sum_{k=2}^{s-1}R_{A^c}^{(k)}(B)
\label{22.20}
\end{eqnarray}
with $R_{A^c}^{(k)}(B)=R_{A^c}^{(k)}(B,\dots,B)\,$, where $R_{A^c}^{(k)}
(B_1,\dots,B_k)$ is a multilinear functional of $B_1,\dots,B_k\in
\Gamma$ with values in $\Gamma$.
It follows from (\ref{22.18}) and (\ref{22.19}) 
that $A^c+B$ is a critical point for $S$ precisely when $R_{A^c}(B)=0$. 
Each element in $T_{A^c}{\cal C}$ has the form $\frac{d}{dt}\Bigl|_{t=0}
A^c(t)=B'(0)$ where $A^c(t)=A^c+B(t)$ is a smooth curve in ${\cal C}$ with
$B(0)=0$. Then $R_{A^c}(B(t))=0$ for all $t\,$, so
\begin{eqnarray*} 
0=\frac{d}{dt}\biggl|_{t=0}R_{A^c}(B(t))=2D_{A^c}(B'(0))
\end{eqnarray*}
which shows that $T_{A^c}{\cal C}\subseteq\ker(D_{A^c})$.
To show the remaining part of (\ref{22.16}) we note from (\ref{22.20}) that
the differential (i.e. the ``Jacobi matrix'') 
of $R_{A^c}$ at $B=0$ is $2D_{A^c}$.
If we were dealing with a smooth
finite-dimensional situation (i.e. if $R_{A^c}$ was
a smooth map between finite-dimensional manifolds)
then the implicit function theorem would imply that the tangentspace to 
the solution space of $R_{A^c}(B)=0$ at $B=0$ is $\ker(D_{A^c})\,$, 
i.e. $T_{A^c}{\cal C}=\ker(D_{A^c})$.
This argument cannot always be extended to infinite-dimensional situations
since the implicit function theorem cannot always be extended to these
situations. It is reasonable to say that the argument can be extended in the 
``generic'' situation though; for example in Yang-Mills gauge theory it
can be extended when $A^c$ is irreducible, and the set of irreducible gauge
field is dense in ${\cal A}$ (see e.g. \cite[Part IV]{BoossBleecker}).
The argument also extends to Chern-Simons 
gauge theory on $S^3$ and lens spaces.
However, there are special cases where the argument cannot be extended and 
where $T_{A^c}{\cal C}{\ne}\ker(D_{A^c})\,$;
examples of this in Chern-Simons gauge theory
have been discussed for example in \cite{Roz}.

To obtain an infrared-finite
perturbative expansion of $I(\alpha;f,S)$ we must rewrite the expression 
(\ref{22.3}) for $I(\alpha;f,S)$ in such a way that the integration in the
functional integral is restricted to a subspace of $\Gamma$ which does not
contain zero-modes for $D_{A^c}$. In \S4 we will show how this can be done for 
gauge theories (i.e. when ${\cal A}$ is a space of gauge fields and $f$ and
$S$ are gauge invariant) using a version of the Faddeev-Popov 
procedure. The procedure rewrites $I(\alpha;f,S)$ in such a way that the 
integration over $B$ is restricted to $T_{A^c}({\cal G}{\cdot}A^c)^{\perp}\,$,
the orthogonal complement to $T_{A^c}({\cal G}{\cdot}A^c)$ in $\Gamma$.
(The price to be paid for this is that a divergent factor $V({\cal G}_0)\,$,
the volume of the subgroup of topologically trivial gauge transformations,
appears in the overall factor multiplying the functional integral. However,
this factor can be avoided by normalising $I(\alpha;f,S)$ by 
$V({\cal G}_0)$ to begin with). 
When $A^c$ is isolated in ${\cal C}$ modulo gauge transformations it follows
from (\ref{22.16}) that $T_{A^c}({\cal G}{\cdot}A^c)=T_{A^c}{\cal C}
=\ker(D_{A^c})$ (in the generic case), so the integration is over
$T_{A^c}({\cal G}{\cdot}A^c)^{\perp}=\ker(D_{A^c})^{\perp}$ which by
definition contains no zero-modes for $D_{A^c}$. In this case the preceding
approach leads to a perturbative expansion of $I(\alpha;f,S)$ in which
infrared divergences do not arise. (Further details will be given in \S4).

We conclude this section by pointing out that the approach to perturbative
expansion described here extends in a straightforward way to situations where
the functional integration is over more that one field, and to the situation
where Grassmannian (anticommuting) fields are involved. We will exploit this
in \S4, where the gauge-fixed expression obtained for $I(\alpha;f,S)$ involves
additional integrations over anticommuting ``ghost'' fields.

\section{The gauge-theoretic setup}

In this section we describe the gauge-theoretic setup which we will be using
in the rest of this paper.
(The definitions and further details can be
found in \cite{Sch(Top)}, \cite[Part IV]{BoossBleecker}).
The space ${\cal A}$ of gauge fields $A$ is the space of connection
1-forms on a principal fibre bundle $P$ over a compact oriented riemannian
manifold $M$ (spacetime) without boundary. We set $n={\dim}M$. 
The structure group (gauge
group) of $P$ is a compact semisimple Lie group $G\,$; we denote its
Lie algebra by ${\bf g}$.
The bundle $P\times_G{\bf g}$ (where $G$ acts on ${\bf g}$ 
by the adjoint representation) is denoted by $\underline{\bf g}\,$,
and $\Omega^q(M,\underline{\bf g})$ denotes the differential forms of degree
$q$ on $M$ with values in $\underline{\bf g}$.
A riemannian metric on $M$ and invariant inner product in ${\bf g}$ 
determine an inner product $<\cdot,\cdot>_q$ in each $\Omega^q(M,\underline
{\bf g})$. Note that ${\cal A}$ is an affine vectorspace modelled on 
$\Omega^1(M,\underline{\bf g})$ (so $\Omega^1(M,\underline{\bf g})$ is the
space $\Gamma$ of \S2).
We will think of ${\cal A}$ as an infinite-dimensional manifold; the
tangentspace at each $A\in{\cal A}$ is $T_{A}{\cal A}=\Omega^1(M,
\underline{\bf g})$. The inner product $<\cdot,\cdot>_1$ in $\Omega^1(M,
\underline{\bf g})$ therefore determines a metric in ${\cal A}\,$,
which formally determines a volume form ${\cal D\/}A$ on ${\cal A}$
(up to a sign).
The curvature (force tensor) $F^A
=dA+\frac{1}{2}{\lbrack}A,A{\rbrack}$ of each $A\in
{\cal A}$ can be considered as an element in $\Omega^2(M,\underline
{\bf g})$. 

The group ${\cal G}$ of gauge transformations (an 
infinite-dimensional Lie group) can be identified with 
$C^{\infty}(M,\underline{G})\,$, the smooth maps from $M$ to
the bundle $\underline{G}=P\times_GG$ (where $G$ acts on itself by the
adjoint action) which map each $x{\in}M$ to the fibre $\underline{G}_x$
above $x$. It acts on ${\cal A}$ and $\Omega^q(M,\underline{\bf g})$
and we denote the action of $\phi\in{\cal G}$ on $A\in{\cal A}$ and
$B\in\Omega^q(M,\underline{\bf g})$ by $\phi{\cdot}A$ and $\phi{\cdot}B$
respectively.
Given a trivialisation of $P$ over a coordinate patch $U{\subseteq}M$
with coordinates $(x^{\mu})$ and given a basis $\{\lambda_j\}$ for
${\bf g}$ we can express $A\in{\cal A}$ and $B\in\Omega(M,\underline{\bf g})$
in the familiar way:
\begin{eqnarray*}
A(x)\Bigl|_U&=&A_{\mu}^i(x)\lambda_idx^{\mu} \\
B(x)\Bigl|_U&=&\frac{1}{q!}B_{\mu_1\dots\mu_q}^i(x)\lambda_idx^{\mu_1}
\wedge\cdots{\wedge}dx^{\mu_q}
\end{eqnarray*}
The trivialisation allows the restriction of each $\phi\in{\cal G}$
to $U$ to be considered as a function from $U$ to $G\,$, and for all
$x{\in}U$ we have the familiar expressions
\begin{eqnarray*}
(\phi{\cdot}A)(x)&=&\phi(x)A(x)\phi^{-1}(x)+\phi(x)d\phi^{-1}(x) \\
&=&\Bigl(\,A_{\mu}^i(x)\phi(x)\lambda_i\phi^{-1}(x)+\phi(x)\partial_{\mu}
\phi^{-1}(x)\Bigr)dx^{\mu} \\
(\phi{\cdot}B)(x)&=&\phi(x)B(x)\phi^{-1}(x) \\
&=&\frac{1}{q!}B_{\mu_1\cdots\mu_q}^i(x)\phi(x)\lambda_i\phi^{-1}(x)
\lambda_idx^{\mu_1}\wedge\cdots{\wedge}dx^{\mu_q}\,.
\end{eqnarray*}
Note for $A\in{\cal A}\;,\;B\in\Omega^1(M,\underline{\bf g})$ and
$\phi\in{\cal G}$ that $A+B\in{\cal A}$ and
\begin{eqnarray}
\phi\cdot(A+B)=\phi{\cdot}A+\phi{\cdot}B
\label{2.a}
\end{eqnarray}

The inner product in each $\Omega^q(M,\underline{\bf g})$ 
is invariant under ${\cal G}$
so the metric 
and volume form ${\cal D\/}A$ 
on ${\cal A}$ are formally invariant under ${\cal G}$. 
The Lie algebra of ${\cal G}$ is $\mbox{Lie}({\cal G})=T_1{\cal G}=
\Omega^0(M,\underline{\bf g})$. The inner product in $T_1{\cal G}=
\Omega^0(M,\underline{\bf g})$ determines a metric on ${\cal G}$ via
the action of ${\cal G}$ on itself; this formally determines a 
${\cal G}$-biinvariant volume form ${\cal D\/}\phi$ on
${\cal G}$ (up to a sign). Note that the subgroup ${\cal G}_0$ of
${\cal G}$ consisting of the topologically trivial gauge transformations
(i.e. the gauge transformations which can be continuously deformed to the 
identity) has the same Lie algebra as ${\cal G}$, i.e. 
$\mbox{Lie}({\cal G}_0)=\mbox{Lie}({\cal G})=\Omega^0(M,\underline{\bf g})$.

The Lie bracket $\lbrack\cdot,\cdot\rbrack$ in ${\bf g}$ determines a graded
product in the space $\Lambda(T_xM)^*\otimes{\bf g}=\oplus_{q=0}^n
\Lambda^q(T_xM)^*\otimes{\bf g}$ for each $x{\in}M\,$, defined by
$\lbrack\omega_x{\otimes}a,\tau_x{\otimes}b\rbrack=\omega_x\wedge\tau_x\otimes
{\lbrack}a,b\rbrack\,$; this determines a product $\lbrack\cdot,\cdot\rbrack$
in the space $\Omega(M,\underline{\bf g})=\oplus_{q=0}^n\Omega^q(M,
\underline{\bf g})$ making $\Omega(M,\underline{\bf g})$ a graded Lie 
algebra.
Each $A\in{\cal A}$ determines covariant derivatives 
$d_q^A:\Omega^q(M,\underline{\bf g})\to\Omega^{q+1}(M,\underline{\bf g}),
\ q=0,1,\dots,n$, with the covariance property
\begin{eqnarray}
d_q^{\phi{\cdot}A}(\phi{\cdot}B)&=&\phi\cdot(d_q^AB)\ \ \ \ \forall\phi\in
{\cal G},\;B\in\Omega^q(M,\underline{\bf g}) \label{2.1} \\
(d_q^{\phi{\cdot}A})^*(\phi{\cdot}B)&=&\phi\cdot((d_q^A)^*B)\ \ \ \forall
\phi\in{\cal G},\;B\in\Omega^{q+1}(M,\underline{\bf g}) \label{2.2} 
\end{eqnarray}
(where $(d_q^A)^*$ is the adjoint of $d_q^A\,$) and the property
\begin{eqnarray}
d_q^{A+B}C=d_q^AC+{\lbrack}B,C\rbrack \ \ \ {\forall}B\in\Omega^1(M,\underline
{\bf g}),\;C\in\Omega^q(M,\underline{\bf g})\,.
\label{2.3}
\end{eqnarray}
The covariant derivative $d_0^A$ is minus the generator of infinitesimal
gauge transformations of $A\,$: 
For all $v\in\mbox{Lie}({\cal G})=\Omega^0(M,\underline{\bf g})$ we have
\begin{eqnarray}
v{\cdot}A:=\frac{d}{dt}\Bigl|_{t=0}\exp(tv){\cdot}A=-d_0^Av
\label{2.4}
\end{eqnarray}

Some notations: If $L$ is a linear map we denote the image and nullspace 
of $L$ by $\mbox{Im}(L)$ and $\ker(L)$ respectively. If $L\,:\,V{\to}V$ is
selfadjoint w.r.t. an inner product in the vectorspace $V$ then $L$
restricts to an invertible map on $\ker(L)^{\perp}\,$ (the orthogonal
complement to $\ker(L)$ in $V$) which we denote by $\tilde{L}\,$, i.e.
\begin{eqnarray}
\tilde{L}:=L\Bigl|_{\ker(L)^{\perp}}\,:\,\ker(L)^{\perp}\stackrel{\cong}
{\longrightarrow}\ker(L)^{\perp}\,.
\label{2.4a}
\end{eqnarray}

We will be using the following general formulae: Let $M_1$ and $M_2$ be 
riemannian manifolds with volume forms ${\cal D}x$ and ${\cal D}y$
respectively, and let $\Phi:M_1{\to}M_2$ be a smooth invertible map.
The differential of $\Phi$ gives invertible linear maps (the 
``Jacobi matrix'')
\begin{eqnarray*}
{\cal D}_x\Phi:T_xM_1{\to}T_yM_2 \ ,\ \ y=\Phi(x)
\end{eqnarray*}
for each $x{\in}M_1$. The inner products in $T_xM_1$ and $T_yM_2$
(given by the riemannian metrics) determine $|\det({\cal D}_x\Phi)|=
\det(({\cal D}_x\Phi)^*{\cal D}_x\Phi)^{1/2}$ (the Jacobi determinant), and
for arbitrary function $h(y)$ on $M_2$ we have the change of variables
formula
\begin{eqnarray}
\int_{M_2}{\cal D\/}y\,h(y)=\int_{M_1}{\cal D\/}x\,\det(({\cal D}_x\Phi)^*
{\cal D\/}_x\Phi)^{1/2}h(\Phi(x))\,.
\label{2.5}
\end{eqnarray}
When $M_2$ is a vectorspace we define the delta-function $\delta(y)$
on $M_2$ by $\int_{M_2}{\cal D\/}y\,h(y)\delta(y)=h(0)$. Then for 
arbitrary function $g(x)$ on $M_1$ we apply (\ref{2.5}) to get the formula
\begin{eqnarray}
\int_{M_1}{\cal D\/}x\,g(x)\delta(\Phi(x))=
\det\Bigl(({\cal D\/}_{\Phi^{-1}(0)}\Phi)^*{\cal D\/}_{\Phi^{-1}(0)}\Phi
\Bigr)^{-1/2}\,g(\Phi^{-1}(0))\,.
\label{2.6}
\end{eqnarray}

\section{Gauge fixing}

In this section we carry out a gauge-fixing of
the normalised functional integral
\begin{eqnarray}
I(\alpha;f,S)=\frac{1}{V({\cal G}_0)}
\int_{\cal A}{\cal D\/}A\,f(A)e^{-\frac{1}{\alpha^2}S(A)}
\label{4.0}
\end{eqnarray}
to formally rewrite it  
in such a way that an infrared-finite perturbative expansion can be obtained
via the approach described in \S2 when the gauge invariant functionals
$f$ and $S$ satisfy the condition (\ref{22.1}). 
The normalisation factor $V({\cal G}_0)$ is the volume of ${\cal G}_0\,$,
a formal, divergent quantity.
The gauge-fixing is a version of the 
Faddeev-Popov procedure with the covariant background field
gauge-fixing condition 
\begin{eqnarray}
(d_0^{A^c})^*(A-A^c)=0
\label{4.1}
\end{eqnarray}
where the background gauge field
$A^c$ is a critical point for $S$ as in (\ref{1.2}). The perturbative 
expansion of (\ref{4.0}) that we obtain will be infrared-finite when $A^c$
is isolated modulo gauge transformations, and we will briefly discuss
the possibility of extending our approach to obtain an infrared-finite
expansion in the general case.
The Faddeev-Popov functional associated with this gauge-fixing condition is
\begin{eqnarray}
P_{A^c}(A)=\int_{{\cal G}_0}{\cal D\/}\phi\,\delta\Bigl(\,(d_0^{A^c})^*
(\phi{\cdot}A-A^c)\Bigr)\,.
\label{4.2}
\end{eqnarray}
Following \cite{Rouet} we have taken the domain of the formal integration
to be the subgroup ${\cal G}_0$ of topologically 
trivial gauge transformations rather than the complete group ${\cal G}$. 
This has the following consequences: (i) To carry out the gauge-fixing
procedure the functionals $f$ and $S$ need only be invariant under
${\cal G}_0$ (rather than ${\cal G}\,$). (ii) To formally evaluate 
(\ref{4.2}) we only need the solutions $\phi$ to 
$(d_0^{A^c})^*(\phi{\cdot}A-A^c)=0$ which belong to ${\cal G}_0$ (rather than
the complete set of solutions in ${\cal G}\,$). As we will see, a consequence
of (ii) is that under a certain assumption (stated below) problems due to
gauge-fixing ambiguities which arise in the usual approach are avoided.
Note that the formal functional (\ref{4.2}) is invariant under ${\cal G}_0$
since the formal measure ${\cal D\/}\phi$ is invariant under ${\cal G}_0$.
Inserting $1=P_{A^c}(A)\,/\,P_{A^c}(A)$ into the integrand in the functional
integral (\ref{4.0}) leads to
\begin{eqnarray}
I(\alpha;f,S)
&=&\frac{1}{V({\cal G}_0)} 
\int_{{\cal G}_0}{\cal D\/}\phi\,\int_{\cal A}{\cal D\/}A\,f(A)
e^{-\frac{1}{\alpha^2}S(A)}P_{A^c}(A)^{-1}\,\delta((d_0^{A^c})^*
(\phi{\cdot}A-A^c)) \nonumber \\
&=&\frac{1}{V({\cal G}_0)}
V({\cal G}_0)\int_{\cal A}{\cal D\/}A\,f(A)e^{-\frac{1}{\alpha^2}S(A)}
P_{A^c}(A)^{-1}\,\delta((d_0^{A^c})^*(A-A^c)) \nonumber \\
&=&\int_{\Omega^1(M,\underline{\bf g})}{\cal D\/}B\,
f(A^c+B)e^{-\frac{1}{\alpha^2}S(A^c+B)}P_{A^c}(A^c+B)^{-1}
\,\delta((d_0^{A^c})^*B) \nonumber \\
& &\label{4.3}
\end{eqnarray}
To obtain the second line we have used the ${\cal G}_0$-invariance
of $f$, $S$ and $P_{A^c}$.
In the last line we have changed variables from
$A\in{\cal A}$ to $B=A-A^c\in\Omega^1(M,\underline{\bf g})\,$; ${\cal D\/}B$
denotes the formal volume form on $\Omega^1(M,\underline{\bf g})$ 
formally determined (up to a sign) by the inner product $<\cdot,\cdot>_1\,$.
Decomposing\footnote{This is really the decomposition of the closure of
$\Omega^1(M,\underline{\bf g})$ w.r.t. $<\cdot,\cdot>_1\,$, but we ignore
technicalities of this kind here and in the following.} 
$\Omega^1(M,\underline{\bf g})=\ker((d_0^{A^c})^*)\oplus\ker((d_0^{A^c})^*)^
{\perp}\ ,\ B=(\tilde{B},C)\ ,\ {\cal D\/}B={\cal D\/}\tilde{B}{\cal D}C\,$
and noting that $\ker((d_0^{A^c})^*)=\mbox{Im}(d_0^{A^c})^{\perp}\,$ 
we use the formula (\ref{2.6}) to integrate over $\ker((d_0^{A^c})^*)^{\perp}$ 
in (\ref{4.3}) and get
\begin{eqnarray}
\lefteqn{I(\alpha;f,S)} \nonumber \\
&=&\det(\tilde{\Delta}_0^{A^c})^{-1/2}
\,\int_{\mbox{Im}(d_0^{A^c})^
{\perp}}{\cal D\/}\tilde{B}\,f(A^c+\tilde{B})e^{\frac{-1}{\alpha^2}
S(A^c+\tilde{B})}P_{A^c}(A^c+\tilde{B})^{-1} \nonumber \\
& &\label{4.4}
\end{eqnarray}
where $\Delta_0^{A^c}=(d_0^{A^c})^*d_0^{A^c}$ and
$\tilde{\Delta}_0^{A^c}$ is the restriction to $\ker(\Delta_0^{A^c})^{\perp}
=\ker(d_0^{A^c})^{\perp}$ as in (\ref{2.4a}); the determinant will be
regularised by zeta-regularisation as discussed below.

The next step is to formally evaluate the Faddeev-Popov functional
\begin{eqnarray}
P_{A^c}(A^c+\tilde{B})=\int_{{\cal G}_0}{\cal D\/}\phi\,\delta\Bigl(\,
(d_0^{A^c})^*(\phi\cdot(A^c+\tilde{B})-A^c)\Bigr)
\label{4.5}
\end{eqnarray}
appearing in (\ref{4.4}) with $\tilde{B}\in\ker((d_0^{A^c})^*)$. 
To do this we must determine the solutions $\phi\in{\cal G}_0$ to 
\begin{eqnarray}
(d_0^{A^c})^*(\,\phi\cdot(A^c+\tilde{B})-A^c)=0
\label{4.6}
\end{eqnarray}
since it is only for these that the integrand in (\ref{4.5}) is non-vanishing.
We see immediately that $\phi=1$ (the identity) is a solution because
$\tilde{B}\in\ker((d_0^{A^c})^*)$.
We now show that each $\phi{\in}H_{A^c}$ is a solution to (\ref{4.6}) where
\begin{eqnarray}
H_{A^c}=\{\,\phi\in{\cal G}_0\;|\;\phi{\cdot}A^c=A^c\,\}\,.
\label{4.7}
\end{eqnarray}
It suffices to show
\begin{eqnarray}
(d_0^{A^c})^*(A-A^c)=0\ \ \Rightarrow
\ \ (d_0^{A^c})^*(\phi{\cdot}A-A^c)=0\ \ \ \ 
\forall\;\phi{\in}H_{A^c}\,.
\label{4.8}
\end{eqnarray}
i.e. the gauge-fixing condition (\ref{4.1}) has ambiguities coming 
from $H_{A^c}$. (The group $H_{A^c}$ is finite-dimensional and can
be identified with a subgroup of $G\,$, see e.g. \cite[p.111-112]{Fuchs}
and the references given there). 
Using (\ref{2.a}) and (\ref{2.2}) we see that for $\phi{\in}H_{A^c}\;$
\begin{eqnarray*}
(d_0^{A^c})^*(\phi{\cdot}A-A^c)&=&(d_0^{A^c})^*(\phi\cdot(A-A^c))=
\phi^{-1}\cdot(d_0^{\phi^{-1}{\cdot}A^c})^*(A-A^c) \\
&=&\phi^{-1}\cdot(d_0^{A^c})^*(A-A^c) 
\end{eqnarray*}
from which (\ref{4.8}) follows.
This shows that $H_{A^c}$ is contained in the solution set to (\ref{4.6});
we now make the following assumption which implies that $H_{A^c}$ is the
complete set of solutions to (\ref{4.6}).

\vspace{1ex}

\noindent {\it \underline{Assumption}: 
If $\phi\in{\cal G}_0$ and $A\in{\cal A}$ 
satisfy $(d_0^{A^c})^*(A-A^c)=0$ and $(d_0^{A^c})^*(\phi{\cdot}A-A^c)=0$
then $\phi{\in}H_{A^c}\,$.}

\vspace{1ex}

\noindent In other words we are assuming that all gauge-fixing ambiguities
in the gauge-fixing condition (\ref{4.1}) come either from gauge 
transformations in $H_{A^c}$ or from {\it topologically non-trivial}
gauge transformations. 
We are not able to prove the assumption but it seems to be compatible
with what is already known about gauge-fixing ambiguities:
The existence of 
gauge-fixing ambiguities was first pointed out by Gribov \cite{Gribov}
who considered the Coulomb gauge-fixing condition with spacetime
$M=S^3\times{\bf R}$ (and gauge fields on a trivial principal fibre bundle).
He showed that there is a collection of gauge transformations
$\{\phi_n\}\;,\ \;n\in{\bf Z}\,$, such that each $\phi_n{\cdot}A$ satisfies the
Coulomb condition when $A=0$. However, it was subsequently shown in 
\cite{Jackiw(PRD)} that all of these gauge transformations are topologically
non-trivial (except the identity transformation $\phi_0=1\,$). 
Therefore, if the general features of gauge ambiguities are the same for 
different gauge-fixing conditions, then Gribov's example of gauge ambiguities
is compatible with our assumption. As far as we are aware there are no 
examples of gauge ambiguities which contradict the assumption.
Also, there does not appear to be any immediate contradiction
between the assumption and the work of I.~Singer \cite{Singer} 
and M.~Narasimhan and T.~Ramadas \cite{Nara}
on the unavoidability of gauge-fixing ambiguities, since this work did not
determine whether the ambiguities came from topologically trivial- or
non-trivial gauge transformations.

In any case our evaluation of the Faddeev-Popov functional (\ref{4.5}),
which takes into account the gauge-fixing ambiguities coming from $H_{A^c}\,$,
is a refinement of the usual evaluation which assumes that there are no
gauge ambiguities. To evaluate (\ref{4.5}) we parameterise a neighbourhood
of $H_{A^c}$ (the solutions to (\ref{4.6})) in ${\cal G}_0$ by two 
coordinates; one of these parameterises directions along $H_{A^c}\,$,
the other parameterises directions transverse to $H_{A^c}$.
The idea is to integrate out the delta-function along the transverse
coordinate and then integrate over $H_{A^c}$.
We decompose $\mbox{Lie}({\cal G}_0)=\Omega^0(M,\underline{\bf g})$ as
$\mbox{Lie}({\cal G}_0)=\mbox{Lie}(H_{A^c})\oplus\mbox{Lie}(H_{A^c})^{\perp}$
and define the map
\begin{eqnarray}
Q\,:\,\mbox{Lie}(H_{A^c})^{\perp}{\times}H_{A^c}\,\to\,{\cal G}_0\ \ \ \ \ \ \ 
Q(v,h):=\exp(v)h
\label{4.9}
\end{eqnarray}
illustrated in the figure below:

\vspace{1ex}

[{\it The figure is not included; it is available on request from the author.}]

\vspace{1ex}

\noindent We will formally show that this map parameterises a neighbourhood of 
$H_{A^c}$ in ${\cal G}_0$ by showing that it is non-degenerate at
$\{0\}{\times}H_{A^c}\,$, i.e. that the ``Jacobi matrix'' of $Q$ at $(0,h)\,$,
\begin{eqnarray}
{\cal D\/}_{(0,h)}Q\,:\,\mbox{Lie}(H_{A^c})^{\perp}{\oplus}T_hH_{A^c}\,
\to\,T_h{\cal G}_0
\label{4.10}
\end{eqnarray}
has non-zero determinant for all $h{\in}H_{A^c}$. In fact we will show that
(\ref{4.10}) is an isometry, from which it follows that
\begin{eqnarray}
|\det({\cal D\/}_{(0,h)}Q)|=1\ \ \ \ \ \ \ \ \ \ \ \ \forall\,h{\in}H_{A^c}\,.
\label{4.11}
\end{eqnarray}
For fixed $h{\in}H_{A^c}$ consider the composition of maps
\begin{eqnarray}
\mbox{Lie}({\cal G}_0)&=&\mbox{Lie}(H_{A^c})^{\perp}\oplus\mbox{Lie}(H_{A^c})
\stackrel{\cong}{\longrightarrow}\mbox{Lie}(H_{A^c})^{\perp}{\oplus}
T_hH_{A^c}\stackrel{{\cal D\/}_{(0,h)}Q}{\longrightarrow}T_h{\cal G}_0
\stackrel{\cong}{\longrightarrow}\mbox{Lie}({\cal G}_0) \nonumber \\
& &\label{4.12}
\end{eqnarray}
where the first map is the isometry given by $(w,a)\mapsto
(w,\frac{d}{dt}\Bigl|_
{t=0}e^{ta}h)$ and the last map is the inverse of the isometry
$\mbox{Lie}({\cal G}_0)\stackrel{\cong}{\to}T_h{\cal G}_0$ given by
$v\mapsto\frac{d}{dt}\Bigl|_{t=0}e^{tv}h\,$. We will show that the composition
of maps (\ref{4.12}) is the identity map on $\mbox{Lie}({\cal G}_0)\,$; 
it then follows that ${\cal D\/}_{(0,h)}Q$ must be an isometry since all 
the other maps in (\ref{4.12}) are isometries. The image of $v\in\mbox{Lie}
({\cal G}_0)$ under the maps in (\ref{4.7}) is
\begin{eqnarray*}
v=(w,a)&{\mapsto}&\frac{d}{dt}\Bigl|_{t=0}(tw\,,e^{ta}h)\mapsto\frac{d}{dt}
\Bigl|_{t=0}Q(tw\,,\,e^{ta}h)=\frac{d}{dt}\Bigl|_{t=0}e^{tw}e^{ta}h \\
&{\mapsto}&\frac{d}{dt}\Bigl|_{t=0}e^{tw}e^{ta}=w+a=v 
\end{eqnarray*}
so (\ref{4.12}) is the identity as claimed.

We now choose a sufficiently small neighbourhood ${\cal N}$ of $\{0\}$ in
$\mbox{Lie}(H_{A^c})^{\perp}\,$, as illustrated in the figure above, 
so that the parameterisation map $Q$ restricts to an invertible map from 
${\cal N}{\times}H_{A^c}$ to a neighbourhood of $H_{A^c}$
in ${\cal G}_0$. (The non-degeneracy of ${\cal D\/}_{(0,h)}Q$ for all
$h{\in}H_{A^c}$ indicates that such a neighbourhood exists; this would 
certainly be the case in a smooth finite-dimensional situation but we have not
proved its existence rigorously in the present infinite-dimensional situation).
Then, since the integrand in the Faddeev-Popov functional
(\ref{4.5}) vanishes outside of
$H_{A^c}$ (by our assumption that $H_{A^c}$ is the complete solution set
to (\ref{4.6})) we can use (\ref{2.5}) to write (\ref{4.5}) as
\begin{eqnarray}
P_{A^c}(A^c+\tilde{B})&=&
\int_{H_{A^c}{\times}{\cal N}}
{\cal D\/}h{\cal D\/}v\,|\det({\cal D\/}_{(v,h)}Q)|
\,\delta\Bigl(\,(d_0^{A^c})^*(e^vh\cdot(A^c+\tilde{B})-A^c)\Bigr)\,.\\
& &\label{4.13}
\end{eqnarray}
From (\ref{2.4}) we see that $\mbox{Lie}(H_{A^c})=\ker(d_0^{A^c})\,$, and
therefore $\mbox{Lie}(H_{A^c})^{\perp}=\ker(d_0^{A^c})^{\perp}=
\mbox{Im}(d_0^{A^c})^*$. For fixed $h{\in}H_{A^c}$ using (\ref{2.4}) again
we see that the Jacobi matrix of the map
\begin{eqnarray*}
v\mapsto(d_0^{A^c})^*(\,e^vh\cdot(A^c+\tilde{B})-A^c)\ \ \ \ \ \ \ 
v\in\mbox{Lie}(H_{A^c})^{\perp}=\ker(d_0^{A^c})^{\perp}
\end{eqnarray*}
at $v=0$ is
\begin{eqnarray}
-(d_0^{A^c})^*d_0^{h\cdot(A^c+\tilde{B})}\biggl|_{\ker(d_0^{A^c})^{\perp}}
\,:\,\ker(d_0^{A^c})^{\perp}\to\ker(d_0^{A^c})^{\perp}\,.
\label{4.13.5}
\end{eqnarray}
Using this together with (\ref{2.6}) and (\ref{4.11}) we integrate out the 
variable $v{\in}{\cal N}\subseteq\mbox{Lie}(H_{A^c})^{\perp}$ in (\ref{4.13})
to get
\begin{eqnarray}
P_{A^c}(A^c+\tilde{B})=\int_{H_{A^c}}{\cal D\/}h\,\Bigl|\det\biggl(\,
(d_0^{A^c})^*d_0^{h\cdot(A^c+\tilde{B})}\biggl|_{\ker(d_0^{A^c})^{\perp}}
\,\biggr)\,\Bigl|^{-1}\;.
\label{4.14}
\end{eqnarray}
From (\ref{2.1}) and (\ref{2.2}) we see that for $h{\in}H_{A^c}\,$,
\begin{eqnarray}
(d_0^{A^c})^*d_0^{h\cdot(A^c+\tilde{B})}&=&(d_0^{A^c})^*\,(h\cdot)\,d_0^
{A^c+\tilde{B}}\,(h^{-1}\cdot)=(h\cdot)\,(d_0^{h^{-1}{\cdot}A^c})^*d_0^
{A^c+\tilde{B}}\,(h\cdot)^{-1} \nonumber \\
&=&(h\cdot)\,(d_0^{A^c})^*\,d_0^{A^c+\tilde{B}}\,(h\cdot)^{-1}\,.
\label{4.14.5}
\end{eqnarray}
The action of $h{\in}H_{A^c}$ on $\Omega^0(M,
\underline{\bf g})$ leaves $\ker(d_0^{A^c})^{\perp}$ invariant
(since the action is by isometries and leaves $\ker(d_0^{A^c})=\mbox{Lie}
(H_{A^c})$ invariant)
and it follows from (\ref{4.14.5}) that
\begin{eqnarray}
\det\Bigl(\,(d_0^{A^c})^*d_0^{h\cdot(A^c+\tilde{B})}\biggl|_{\ker(d_0^{A^c})
^{\perp}}\Bigr)=\det\Bigl(\,(d_0^{A^c})^*d_0^{A^c}\biggl|_{\ker(d_0^{A^c})
^{\perp}}\Bigr)
\label{4.14.7}
\end{eqnarray}
independent of $h$. Substituting this into (\ref{4.14}) leads to
\begin{eqnarray}
P_{A^c}(A^c+\tilde{B})=V(H_{A^c})\,\Bigl|\det\biggl(\,(d_0^{A^c})^*
d_0^{A^c+\tilde{B}}\biggl|_{\ker(d_0^{A^c})^{\perp}}\,\biggr)\Bigl|^{-1}
\label{4.15}
\end{eqnarray}
where $V(H_{A^c})$ is the volume\footnote{If $H_{A^c}$ is discrete (i.e.
if $A^c$ is weakly irreducible) then $V(H_{A^c})$ is replaced by the
number $|H_{A^c}|$ of elements in $H_{A^c}$.} 
of $H_{A^c}\subset{\cal G}_0\,$.
Our calculation above contains an implicit assumption that the map
(\ref{4.13.5}) is non-degenerate (this is a requirement for using the 
formula (\ref{2.6})). 
This assumption is less crucial than our previous one for the following
reason. For small $\alpha$ the functional integral (\ref{4.4}) is (formally)
dominated by the contribution from a neighbourhood of 0 in 
$\mbox{Im}(d_0^{A^c})^{\perp}\,$, 
and the map (\ref{4.13.5}) is non-degenerate for
such a neighbourhood (provided that it is sufficiently small) since
$(d_0^{A^c})^*d_0^{A^c}$ is non-degenerate on $\ker(d_0^{A^c})^{\perp}$.
With this assumption we can state at the formal
level that the determinant in (\ref{4.15}) is non-zero and has the same
sign for all $\tilde{B}$.
Since $(d_0^{A^c})^*d_0^{A^c}$ is a strictly positive map on $\ker
(d_0^{A^c})^{\perp}$ the sign of the determinant is positive and we can
discard the numerical signs in (\ref{4.15}). 
This leads to the final result:
\begin{eqnarray}
P_{A^c}(A^c+\tilde{B})=V(H_{A^c})\,\det\Bigl(\,(d_0^{A^c})^*d_0^{A^c+\tilde{B}}
\biggl|_{\ker(d_0^{A^c})^{\perp}}\,\Bigr)^{-1}\,.
\label{4.16}
\end{eqnarray}
This expression differs from the one obtained from the usual evaluation of the
Faddeev-Popov functional (which does not take into account the gauge 
ambiguities coming from $H_{A^c}\,$): The volume factor $V(H_{A^c})$ appears,
and the map $(d_0^{A^c})^*d_0^{A^c+\tilde{B}}$ 
in the determinant is restricted to 
$\ker(d_0^{A^c})^{\perp}$. This latter feature is crucial for
avoiding infrared divergences in the ghost propagator, as we will see below.
The volume factor $V(H_{A^c})$ is crucial for the metric-independence of the
overall term multiplying the perturbation series for the partition function
in Chern-Simons gauge theory, as we will see in \S5 ((\ref{7.5}) and the 
subsequent discussion), and for reproducing the large $k$ limits of
non-perturbative expressions for the Chern-Simons partition function
obtained from the prescription of \cite{W(Jones)} (cf. the discussion in the
conclusion).

Substituting (\ref{4.16}) into the expression (\ref{4.4}) 
for the functional integral leads to a factor
$\det\Bigl(\,(d_0^{A^c})^*d_0^{A^c}\Bigl|_{\ker(d_0^{A^c})^{\perp}}\Bigr)$
in the integrand. Following the Faddeev-Popov procedure we write this 
determinant as a formal Grassmann integral over independent anticommuting
variables (``ghost fields'')
$\bar{C}\,,\,C\in\ker(d_0^{A^c})^{\perp}\,$:
\begin{eqnarray}
\det\Bigl(\,(d_0^{A^c})^*d_0^{A^c+\tilde{B}}\biggl|_{\ker(d_0^{A^c})^{\perp}}
\,\Bigr)&=&\int_{\ker(d_0^{A^c})^{\perp}\oplus\ker(d_0^{A^c})^{\perp}}
{\cal D\/}\bar{C}{\cal D\/}C\,e^{-<\bar{C}\,,\,(d_0^{A^c})^*d_0^{A^c+\tilde{B}}
C>_0}\,. \nonumber \\
& &\label{4.17}
\end{eqnarray}
Using (\ref{2.3}) the term in the exponential in the integrand can be 
written as
\begin{eqnarray}
<\bar{C}\,,\,(d_0^{A^c})^*d_0^{A^c+\tilde{B}}C>_0\,=\,
<\bar{C}\,,\,\Delta_0^{A^c}
C>_0+<\bar{C}\,,\,(d_0^{A^c})^*\lbrack\tilde{B}\,,\,C\rbrack>_0
\label{4.17a}
\end{eqnarray}
Substituting (\ref{4.16}) into the expression (\ref{4.4}) for the functional
integral and writing the determinant as in (\ref{4.17})--(\ref{4.17a}) 
we finally arrive
at the gauge-fixed expression for $I(\alpha;f,S)\,$:
\begin{eqnarray}
\lefteqn{I(\alpha;f,S)_{{\lbrack}A^c\rbrack}} \nonumber \\
&=&V(H_{A^c})^{-1}\det(\tilde{\Delta}_0^{A^c})^{-1/2} 
\int_{\mbox{Im}(d_0^{A^c})^{\perp}\oplus\ker(d_0^{A^c})^{\perp}
\oplus\ker(d_0^{A^c})^{\perp}}{\cal D\/}\tilde{B}{\cal D\/}\bar{C}
{\cal D\/}C\Big\{f(A^c+\tilde{B}) \nonumber \\
& &\times\,\exp\Bigl(\,-\frac{1}{\alpha^2}S(A^c+\tilde{B})
-<\bar{C},\,\Delta_0^{A^c}C>_0
-<\bar{C}\,,\,(d_0^{A^c})^*\lbrack\tilde{B}\,,\,C\rbrack>_0\Bigr)\Big\}
\nonumber \\
& &\label{4.19}
\end{eqnarray}
It is easy to show (using the ${\cal G}_0-$invariance of $f$ and 
$S\,$) that $I(\alpha;f,S)_{{\lbrack}A^c\rbrack}$ is (formally) 
unchanged when $A^c$ is replaced by $\phi{\cdot}A^c$ for any 
$\phi\in{\cal G}_0\,$, i.e. depends only on the orbit ${\lbrack}A^c\rbrack=
{\cal G}_0{\cdot}A^c$ of ${\cal G}_0$ through $A^c\,$ (we leave 
the verification of this to the reader).

The expression (\ref{4.19})
can be perturbatively expanded by the approach described
in \S2, with a straightforward modification to take account of
the fact that the integration over the variable $\tilde{B}$ 
is restricted to $\mbox{Im}(d_0^{A^c})^{\perp}$ and the fact that there
are additional integrations of Grassmannian variables $\bar{C}$ and $C$
over $\ker(d_0^{A^c})^{\perp}$.
The operator $\Delta_0^{A^c}$ in the quadratic term for the ghost variables
in the exponential in (\ref{4.19}) plays an analogous role to $D_{A^c}$
in the perturbative expansion.
From (\ref{2.4}) we see that $\mbox{Im}(d_0^{A^c})
=T_{A^c}({\cal G}{\cdot}A^c)$. It follows from (\ref{22.17}) that 
$\mbox{Im}(d_0^{A^c})^{\perp}=T_{A^c}({\cal G}{\cdot}A^c)^{\perp}
\supseteq\ker(D_{A^c})^{\perp}$ is 
invariant under $D_{A^c}\,$, since $\ker(D_{A^c})^{\perp}$ is invariant under
$D_{A^c}$. We can therefore choose orthonormal bases $\{B_j\}_{j=0,1,2,\dots}$
and $\{C_l\}_{l=0,1,2,\dots}$ for $\mbox{Im}(d_0^{A^c})^{\perp}$ and
$\ker(d_0^{A^c})^{\perp}=\ker(\Delta_0^{A^c})^{\perp}$ 
respectively, consisting of eigenvectors for $D_{A^c}$ and $\Delta_0^{A^c}$
as follows:
\begin{eqnarray}
D_{A^c}B_j&=&\lambda(j)B_j \ \ \ \ \ \ \ \ \ \ \ \ \ \ j=0,1,2,\dots
\label{44.1} \\
0{\le}|\lambda(0)|\le{\dots}&{\le}&|\lambda(j)|{\le}|\lambda(j+1)|\le\dots
\to\infty\ \ \ \mbox{for}\ j\to\infty \label{44.2} \\
\Delta_0^{A^c}C_l&=&\mu(l)C_l\ \ \ \ \ \ \ \ \ \ \ \ \ \ l=0,1,2,\dots
\label{44.3} \\
0<\mu(0)\le{\dots}&{\le}&{\mu}(l)\le\mu(l+1)\le\dots\to\infty\ \ \ \mbox{for}
\ l\to\infty\,. \label{44.4}
\end{eqnarray}
(The eigenvectors and eigenvalues above depend of course on $A^c$ but for
the sake of notational simplicity we suppress this in the notation).
In the case of Yang-Mills- and Chern-Simons gauge theories the fact that
$D_{A^c}$ and $\Delta_0^{A^c}$ have discrete spectra, and the properties
(\ref{44.2}) and (\ref{44.4}) of the eigenvalues, follow via standard 
mathematical results from the relationship that these operators have to the
elliptic complexes (\ref{3.9}) and (\ref{3.13}) below. 
(This result relies on $M$ being compact, riemannian and without boundary).
Carrying out the perturbative expansion of (\ref{4.19}) by the method 
of \S2 leads to
\begin{eqnarray}
\lefteqn{I(\alpha;f,S)_{{\lbrack}A^c\rbrack}} \nonumber \\
&=&V(H_{A^c})^{-1}\,\det\Bigl(\,\frac{1}{\pi\alpha^2}
\tilde{D}_{A^c}\Bigr)^{-1/2}\,\det(\tilde{\Delta}_0^{A^c})^{1/2}\,
e^{-\frac{1}{\alpha^2}S(A^c)} 
\nonumber \\
& &\times\,\Big{\lbrack}f(A^c)+
\sum_{k=1}^{\infty}\alpha^k\biggl(\,\sum_{N=1}^k\,\sum_
{j_1,\dots,j_N}f_{A^c}^{j_1{\cdots}j_N}G_{A^c}^{(N,k-N)}(j_1,\dots,j_N)\,
\biggr)\,\Big{\rbrack}
\label{44.5}
\end{eqnarray}
where $\alpha^kG_{A^c}^{(N,k)}(j_1,\dots,j_N)$ is obtained from Feynman
diagrams as in \S2. There are now additional building blocks for the 
diagrams due to the additional integrations over the Grassmannian variables
$\bar{C}$ and $C$ in (\ref{4.19}). The building blocks for the diagrams in 
this case, and the factors which each of these contribute, are as follows:

\setlength{\unitlength}{0.01250000in}%
\begingroup\makeatletter\ifx\SetFigFont\undefined%
\gdef\SetFigFont#1#2#3#4#5{%
  \reset@font\fontsize{#1}{#2pt}%
  \fontfamily{#3}\fontseries{#4}\fontshape{#5}%
  \selectfont}%
\fi\endgroup%
\begin{picture}(317,215)(73,618)
\thinlines
\put(295,765){\vector( 0, 1){0}}
\put(315,765){\oval( 40, 40)[bl]}
\put(320,770){\line( 0, 1){ 50}}
\put(321,770){\line(-3, 1){ 45}}
\put(320,771){\line( 3, 1){ 49.500}}
\put(321,770){\line( 2,-5){ 18.276}}
\put( 75,775){\line( 1, 0){ 60}}
\put(320,650){\line( 0, 1){ 50}}
\put(340,635){\line(-4, 3){ 20}}
\put(320,650){\vector(-4,-3){ 20}}
\put(300,635){\line(-4,-3){ 20}}
\put(360,620){\vector(-4, 3){ 20}}
\put( 75,670){\vector( 1, 0){ 30}}
\put(105,670){\line( 1, 0){ 30}}
\put(375,785){\makebox(0,0)[lb]{\smash{\SetFigFont{12}{14.4}{\rmdefault}{\mddefault}{\updefault}$i_2$
}}}
\put(320,825){\makebox(0,0)[lb]{\smash{\SetFigFont{12}{14.4}{\rmdefault}{\mddefault}{\updefault}$i_1$
}}}
\put(345,720){\makebox(0,0)[lb]{\smash{\SetFigFont{12}{14.4}{\rmdefault}{\mddefault}{\updefault}$i_3$
}}}
\put(255,785){\makebox(0,0)[lb]{\smash{\SetFigFont{12}{14.4}{\rmdefault}{\mddefault}{\updefault}$i_p$
}}}
\put(100,675){\makebox(0,0)[lb]{\smash{\SetFigFont{12}{14.4}{\rmdefault}{\mddefault}{\updefault}$l$
}}}
\put( 90,750){\makebox(0,0)[lb]{\smash{\SetFigFont{12}{14.4}{\rmdefault}{\mddefault}{\updefault}$\frac{1}{4\lambda(j)}$
}}}
\put( 90,645){\makebox(0,0)[lb]{\smash{\SetFigFont{12}{14.4}{\rmdefault}{\mddefault}{\updefault}$\frac{1}{\mu(l)}$
}}}
\put(100,780){\makebox(0,0)[lb]{\smash{\SetFigFont{12}{14.4}{\rmdefault}{\mddefault}{\updefault}$j$
}}}
\put(310,700){\makebox(0,0)[lb]{\smash{\SetFigFont{12}{14.4}{\rmdefault}{\mddefault}{\updefault}$i$
}}}
\put(270,630){\makebox(0,0)[lb]{\smash{\SetFigFont{12}{14.4}{\rmdefault}{\mddefault}{\updefault}$l_2$
}}}
\put(360,630){\makebox(0,0)[lb]{\smash{\SetFigFont{12}{14.4}{\rmdefault}{\mddefault}{\updefault}$l_1$
}}}
\put(390,640){\makebox(0,0)[lb]{\smash{\SetFigFont{12}{14.4}{\rmdefault}{\mddefault}{\updefault}$\alpha\tilde{S}_{A^c}^{il_1l_2}$
}}}
\put(390,740){\makebox(0,0)[lb]{\smash{\SetFigFont{12}{14.4}{\rmdefault}{\mddefault}{\updefault}$\alpha^{p-2}S_{A^c}^{i_1{\cdots}i_p}\ ,\;p=3,4,\dots,s$
}}}
\end{picture}

\noindent where $\lambda(j)$ and $\mu(l)$ are as in (\ref{44.1}) and 
(\ref{44.3}) respectively, 
$S_{A^c}^{i_1{\cdots}i_p}=S_{A^c}^{(p)}(B_{i_1},\dots,B_{i_p})$ as in \S2,
and the factor for the new vertex is
$\alpha\tilde{S}_{A^c}^{il_1l_2}=\alpha<\bar{C}_{l_1},
(d_0^{A^c})^*{\lbrack}B_i,C_{l_2}\rbrack>_0$.
Each diagram for $\alpha^kG_{A^c}^{(N,k)}(j_1,\dots,j_N)$ has $N$ external
unoriented lines (``gauge lines'') as before, however the diagrams may
now contain internal oriented lines (``ghost lines'') and the
``gauge-ghost'' vertex in addition to the unoriented gauge lines and gauge
vertices. All closed loops formed by the ghost lines must
be oriented. The only change in the rules stated in \S2 for obtaining the
expression corresponding to a given diagram is that a factor $-1$ must be
included for each closed loop formed by the ghost lines.
To explicitly determine the propagators and vertex factors for the Feynman
diagrams the eigenvectors and eigenvalues of $D_{A^c}$ and $\Delta_0^{A^c}$
must be determined. This is a non-trivial problem in general; however in
Chern-Simons gauge theory with spacetime $S^3$ or a lens space techniques
already exist for determining these as we discuss in the conclusion.

The situation with regard to infrared divergences in the expressions for
the Feynman diagrams is as follows. 
Since $\Delta_0^{A^c}$ has no zero-modes in $\ker(\Delta_0^{A^c})^{\perp}$
the propagator $\frac{1}{\mu(l)}$ for the ghost lines is finite for all
$l=0,1,2,\dots$ (cf. ({\ref{44.2}))
and does not give rise to infrared divergences.
For the gauge line propagator $\frac{1}{4\lambda(j)}$ to be finite for all
$j=0,1,2,\dots$ the operator $D_{A^c}$ must have no zero-modes in
$\mbox{Im}(d_0^{A^c})^{\perp}$ (cf. (\ref{44.1})). 
Since $\mbox{Im}(d_0^{A^c})=T_{A^c}({\cal G}{\cdot}A^c)$ we see from
(\ref{22.16}) that this is the case precisely when $A^c$ is isolated
modulo gauge transformations.
Therefore, when $A^c$ is isolated modulo gauge transformations
the expressions for the Feynman diagrams are completely free of 
infrared divergences. 
In this case the operators $\tilde{D}_{A^c}$ and $\tilde{\Delta}_0^{A^c}$
in (\ref{44.5}) are the restrictions of $D_{A^c}$ and $\Delta_0^{A^c}$ to 
invertible maps on
$\ker(D_{A^c})^{\perp}$ and $\ker(\Delta_0^{A^c})^{\perp}$ respectively. 
We show below that for Yang-Mills- and Chern-Simons gauge theories
the determinants in (\ref{44.5}) can be given well-defined meaning via 
zeta-regularisation -this enables the $\alpha$-dependence of 
$\det\Bigl(\frac{1}{\pi\alpha^2}\tilde{D}_{A^c}\Bigr)$ to be extracted.
A straightforward consequence of the ${\cal G}_0$-invariance of $S$ and $f$
is that the expressions for the Feynman diagrams depend only on the orbit
${\lbrack}A^c\rbrack$ of ${\cal G}_0$ through $A^c$. We omit the details,
except to note that as a consequence of the ${\cal G}_0$-invariance of $S$
the operator $D_{A^c}$ has the covariance property
\begin{eqnarray}
D_{\phi{\cdot}A^c}(\phi{\cdot}B)=\phi\cdot(D_{A^c}B)\ \ \ \ \ \ \ 
{\forall}B\in\Omega^1(M,\underline{\bf g})\ ,\ \ \phi\in{\cal G}_0
\label{3.14}
\end{eqnarray}
and the operator $\Delta_0^{A^c}$ has the same covariance property due to 
(\ref{2.2})-(\ref{2.3}).

When $A^c$ is not isolated
modulo gauge transformations the gauge propagator $\frac{1}{4\lambda(j)}$
diverges for sufficiently small $j$ and infrared divergences are present.
It would be very desirable to have a method for perturbative expansion
which is also infrared-finite when $A^c$ is not isolated modulo gauge
transformations, particularly for Yang-Mills gauge theory where the instanton
modulispaces have non-zero dimension in general. We now give a brief, rough
sketch of how it may be possible to achieve this using a version of the
preceding approach to perturbative expansion. 
For small $\alpha$ the functional integral $I(\alpha;f,S)$ is (formally)
dominated by the contributions from neighbourhoods of the critical points
for $S$. We can therefore approximate $I(\alpha;f,S)$ for small $\alpha$ by
\begin{eqnarray}
I(\alpha;f,S)_{{\cal N}({\cal C})}=\frac{1}{V({\cal G}_0)}
\int_{{\cal N}({\cal C})}{\cal D\/}A\,f(A)e^{-\frac{1}{\alpha^2}S(A)}
\label{zz.1}
\end{eqnarray}
where ${\cal N}({\cal C})\subseteq{\cal A}$ is a thin ${\cal G}_0$-invariant
neighbourhood of the space ${\cal C}$ of critical points for $S$. In fact
it is conceivable that the perturbative expansion of
\begin{eqnarray}
\frac{1}{V({\cal G}_0)}\int_{{\cal A}-{\cal N}({\cal C})}{\cal D\/}A\,f(A)
e^{-\frac{1}{\alpha^2}S(A)}
\label{zz.2}
\end{eqnarray}
vanishes (at the formal level); this is claimed (without argument) in
\cite[p.2]{Ax(hep-th)} in the context of Chern-Simons gauge theory.
If this is the case then a perturbative expansion of $I(\alpha;f,S)$
is obtained by perturbatively expanding (\ref{zz.1}). A perturbative 
expansion of (\ref{zz.1}) can be obtained using the techniques developed in
the preceding. We find
\begin{eqnarray}
I(\alpha;f,S)_{{\cal N}({\cal C})}=\int_{{\cal C}/{\cal G}_0}
{\cal D\/}{\lbrack}A^c\rbrack\,
\tilde{I}(\alpha;r_{\cal C}{\cdot}f,S)_{{\lbrack}A^c
\rbrack}
\label{zz.3}
\end{eqnarray}
where $\tilde{I}(\alpha;r_{\cal C}
{\cdot}f,S)_{{\lbrack}A^c\rbrack}$ is given by
(\ref{4.19}) with $f$ replaced by $r_{\cal C}{\cdot}f$ and 
$\mbox{Im}(d_0^{A^c})^{\perp}$ replaced by $\ker(D_{A^c})^{\perp}$
in the integration (so the gauge propagator no longer gives rise to infrared
divergences in the perturbative expansion).
Here $r_{\cal C}$
is a ``measure function'' which we are not able to determine in general.
If the geometry of ${\cal C}$ (induced by the metric in ${\cal A}$) happens
to be flat in the directions orthogonal to the orbits of ${\cal G}_0$ in
${\cal C}$ (or equivalently, if the geometry of ${\cal C}\Big/{\cal G}_0$
is flat) then $r_{\cal C}=1$. In general we can only say that 
$r_{\cal C}(A^c)=1$ for all $A^c\in{\cal C}$.
In arriving at (\ref{zz.3}) we have used a formal generalisation of the 
following observation: For $a\,,\,b\,,\,c>0$ the asymptotics (i.e. 
Taylor expansion) of $\int_a^{\infty}\frac{x^b}{\alpha^c}e^{-\frac{1}
{\alpha^2}x^2}dx$ for $\alpha{\to}0$ vanishes since
\begin{eqnarray}
\frac{d^p}{d\alpha^p}\biggl|_{\alpha=0}\int_a^{\infty}\frac{x^b}{\alpha^c}
e^{-\frac{1}{\alpha^2}x^2}dx=0\ \ \ \ \mbox{for all}\ \ p=0,1,2,\dots
\label{zz.4}
\end{eqnarray}
This is easily shown using the fact that $\frac{1}{y^a}e^{-\lambda}{y^2}
{\to}0$ for $y{\to}0$ for all $a\,,\,\lambda>0$. 
(It seems plausible that (\ref{zz.4}) might also be used to show that the 
asymptotics of (\ref{zz.2}) vanish).

The perturbative expansion of (\ref{zz.1}) is obtained by substituting
(\ref{44.5}) in (\ref{zz.3}) with $f$ replaced by $r_{\cal C}{\cdot}f$.
The higher order terms in the expansion are undetermined since we have not
been able to determine $r_{\cal C}$ in general. However, since 
$r_{\cal C}(A^c)=1$ for $A^c\in{\cal C}$ we obtain an expression for the
lowest order term in the expansion:
\begin{eqnarray}
\int_{{\cal C}/{\cal G}_0}{\cal D\/}{\lbrack}A^c\rbrack\, 
V(H_{A^c})^{-1}\det\Bigl(\,\frac{1}{\pi\alpha^2}
\tilde{D}_{A^c}\Bigr)^{-1/2}\det(\tilde{\Delta}_0^{A^c})^{1/2}
f(A^c)e^{-\frac{1}{\alpha^2}S(A^c)}
\label{zz.5}
\end{eqnarray}
This reproduces the formula \cite[App. II (9)]{Sch(Inst)} for the weak 
coupling limit of $I(\alpha;f,S)$.

We conclude this section by considering two specific gauge theories, the 
Yang-Mills- and Chern-Simons theories, giving expressions for the vertex
factors in the Feynman diagrams for these theories and showing that the 
determinants in (\ref{zz.5}) can be zeta-regularised.
The action functional for Yang-Mills gauge theory on 4-dimensional $M$ is
\begin{eqnarray}
S_{YM}(A)=\frac{1}{2}<F^A\,,F^A>_2
\label{3.3a}
\end{eqnarray}
In this case we take ${\cal C}$ to be the set of absolute minima for $S_{YM}$
(rather than the complete set of critical points for $S_{YM}$). 
The topological number (2nd Chern character) of the principal fibre bundle $P$
is $Q_P=\frac{1}{16\pi^2}<F^A,{\ast}F^A>\,$ (independent of $A\in
{\cal A}\,$); we can assume without loss of generality that $Q_P{\ge}0$ since 
$Q_P$ changes sign when the orientation of $M$ is reversed.
A standard calculation gives
\begin{eqnarray}
S_{YM}(A)=8\pi^2Q_P+<\pi_-F^A,\pi_-F^A>_2
\label{3.4}
\end{eqnarray}
where $\pi_-=\frac{1}{2}(1-\ast)$ on $\Omega^2(M,\underline{\bf g})\,$,
showing that ${\cal C}$ is the space of instantons on $P$, 
i.e. the solutions to $\pi_-F^{A^c}=0$. 
A straightforward calculation using (\ref{3.4}) and
(\ref{2.3}) shows that for $A^c\in{\cal C}\,$,  
\begin{eqnarray}
S_{YM}(A^c+B)&=&8\pi^2Q_P+<B,(\pi_-d_1^{A^c})^*\pi_-d_1^{A^c}B>_1 \nonumber\\
& &\,+<\pi_-d_1^{A^c}B,\pi_-{\lbrack}B,B\rbrack>_2
+\frac{1}{4}<\pi_-{\lbrack}B,B\rbrack,\pi_-{\lbrack}B,B\rbrack>_2 \nonumber \\
& &\label{3.5}
\end{eqnarray}
This shows that $S_{YM}$ satisfies the condition (\ref{22.1}) with $s=4$ and
\begin{eqnarray}
D_{A^c}&=&(\pi_-d_1^{A^c})^*\pi_-d_1^{A^c} \label{3.6} \\
S_{A^c}^{(3)}(B_1,B_2,B_3)&=&<\pi_-d_1^{A^c}B_1,\pi_-{\lbrack}B_2,B_3
\rbrack>_2 \label{3.7} \\
S_{A^c}^{(4)}(B_1,B_2,B_3,B_4)&=&\frac{1}{4}<\pi_-{\lbrack}B_1,B_2\rbrack,
\pi_-{\lbrack}B_3,B_4\rbrack>_2 \label{3.8}
\end{eqnarray}
The gauge vertex factors ${\alpha}S_{A^c}^{i_1i_2i_3}$ 
and $\alpha^2S_{A^c}^{i_1i_2i_3i_4}$
for the Feynman diagrams are obtained from 
(\ref{3.7})--(\ref{3.8}) as described in \S2 (below (\ref{22.6})).
The zeta-regularisability of the determinants in (\ref{zz.5}) 
follows in this case
from the relationships of the operators $D_{A^c}$ and $\Delta_0^{A^c}$ 
to the operators appearing in the elliptic self-dual complex
\begin{eqnarray}
0\longrightarrow\Omega^0(M,\underline{\bf g})\stackrel{d_0^{A^c}}
{\longrightarrow}\Omega^1(M,\underline{\bf g})\stackrel{\pi_-d_1^{A^c}}
{\longrightarrow}\Omega_-^2(M,\underline{\bf g}){\longrightarrow}0
\label{3.9}
\end{eqnarray}
where $\Omega_-^2(M,\underline{\bf g})=\pi_-(\Omega^2(M,\underline{\bf g}))$.
An argument analogous to the one given in \cite{AdSe(PLB)} shows that
the zeta-regularisations of the determinants in (\ref{zz.5}) are well-defined
and lead to
\begin{eqnarray}
\det\Bigl(\frac{1}{\pi\alpha^2}D_{A^c}\Bigr)^{-1/2}
\,\sim\,\vert\alpha\vert^{\zeta(D_{A^c})}
\label{zz.6}
\end{eqnarray}
where $\zeta(D_{A^c})$ is the analytic continuation to 0 of the zeta-function 
for $D_{A^c}$.

The action functional for Chern-Simons  
gauge theory\footnote{We assume for simplicity here that the 
principal fibre bundle $P$ is trivial so that the gauge fields can be 
identified with the ${\bf g}$-valued 1-forms on $M$. This is always
the case when $G=SU(2)\,$; see \cite{Freed} for the general case}  
on 3-dimensional $M$ is
\begin{eqnarray}
-iS_{CS}(A)=-i\frac{1}{4\pi}\int_MTr(A{\wedge}dA+\frac{2}{3}A{\wedge}A
{\wedge}A)
\label{3.3b}
\end{eqnarray}
with the trace taken in the fundamental representation.
The natural parameter for this theory is $k=\frac{1}{\alpha^2}$. The parameter
$k$ is usually required to be integer-valued, since it is only then that
$\exp(ikS_{CS}(A))$ is gauge-invariant. However, $S_{CS}(A)$ is invariant under
the subgroup ${\cal G}_0\,$, and since this is all that is required in our 
method for perturbative expansion $k$ may take arbitrary real values.
The set ${\cal C}$ of critical points for $iS_{CS}$
consists of the flat gauge fields on $P\,$,
i.e. the solutions to $F^{A^c}=0$. For $A^c\in{\cal C}$ a simple calculation 
gives
\begin{eqnarray}
-iS_{CS}(A^c+B)&=&-iS_{CS}(A^c)+<B,i(\frac{1}{4\pi\lambda_{\bf g}}
{\ast}d_1^{A^c})B>_1
-\frac{i}{4\pi}\int_M\mbox{Tr}(B{\wedge}B{\wedge}B) \nonumber \\
& &\label{3.10}
\end{eqnarray}
where $\ast$ is the Hodge star operator and
the invariant inner product in ${\bf g}$ used in constructing the 
inner product $<\cdot,\cdot>_1$ in $\Omega^1(M,\underline{\bf g})$ is
taken to be $<a,b>_{\bf g}=-\lambda_{\bf g}\mbox{Tr}(ab)$ with 
$\lambda_{\bf g}>0$ an arbitrary scaling parameter.
This shows that $-iS_{CS}$ satisfies the
condition (\ref{22.1}) with $s=3$ and
\begin{eqnarray}
D_{A^c}&=&\frac{1}{4\pi\lambda_{\bf g}}{\ast}d_1^{A^c} \label{3.11} \\
-iS_{A^c}^{(3)}(B_1,B_2,B_3)&=&-i\frac{1}{4\pi}\int_M\mbox{Tr}(B_1{\wedge}
B_2{\wedge}B_3)\,. \label{3.12}
\end{eqnarray}
In this case there is one gauge vertex with factor 
$-i\frac{1}{\sqrt{k}}S_{A^c}^{i_1i_2i_3}\,$, obtained from (\ref{3.12})
as described below (\ref{22.6}) in \S2.
The zeta-regularisability of the determinants in (\ref{zz.5}) also
follows in this case
from the relationships of the operators $D_{A^c}$ and $\Delta_0^{A^c}$ 
to an elliptic complex, namely the twisted de Rham complex
\begin{eqnarray}
0\longrightarrow\Omega^0(M,\underline{\bf g})\stackrel{d_0^{A^c}}
{\longrightarrow}\Omega^1(M,\underline{\bf g})\stackrel{d_1^{A^c}}
{\longrightarrow}\Omega^2(M,\underline{\bf g})
\stackrel{d_2^{A^c}}
{\longrightarrow}\Omega^3(M,\underline{\bf g}){\longrightarrow}0
\label{3.13}
\end{eqnarray}
In \cite{AdSe(PLB)} it was shown that the zeta-regularisations of the
determinants in (\ref{zz.5}) are well-defined and, setting 
$k=\frac{1}{\alpha^2}\,$,
\begin{eqnarray}
\det\Bigl(\frac{k}{\pi}iD_{A^c}\Bigr)^{-1/2}
\;{\sim}\;k^{(-{\dim}H^0(d^{A^c})+{\dim}H^1(d^{A^c}))/2}
\end{eqnarray}
where $H^q(d^{A^c})$ is the q'th cohomology space of (\ref{3.13}).
(A more explicit expression is given in (\ref{7.5}) in the following section).

Note that in these examples the requirement $\mbox{Im}(d_0^{A^c})
=\ker(D_{A^c})$ for the absence of infrared divergences in the gauge 
propagator is equivalent to the vanishing of the 1st cohomology space
$H^1(A^c)$ for the complex (\ref{3.9}) or (\ref{3.13}).

Finally, an example of functional $f$ satisfying the condition (\ref{22.1}) 
is the Wilson loop functional,
\begin{eqnarray}
f_{(\gamma\,,\rho)}(A)=Tr\,\Bigl({\cal P}\exp(\oint_{\gamma}\rho(A))\Bigr)
\label{3.3c}
\end{eqnarray}
where $\gamma$ is a closed curve in $M$, $\rho$ is a representation of the 
gauge group $G$ and ${\cal P}$ denotes path-ordering. 
(I.e. $f_{(\gamma,\rho)}(A)$
is the trace of the holonomy of $A$ around $\gamma$ in the representation
$\rho\,$). 
Setting $A=A^c+B$ in (\ref{3.3c}) and expanding the exponential as a power
series it is easy to see that (\ref{22.1}) is satisfied for arbitrary
$A^c$ and $B$.

\section{Perturbative expansion in Chern-Simons gauge theory}

In this section we specialise to 
Chern-Simons gauge theory on 3-dimensional $M$.
We begin by pointing out that the approach to 
perturbative expansion given in the preceding coincides with
the superfield approach of Axelrod and Singer in \cite{AxSi1}
when their condition ($A^c$ acyclic) is satisfied.
We then go on to show that the perturbative expansion of the partition
function is formally metric-independent when $A^c$ is
isolated modulo gauge transformations. This was shown by Axelrod and
Singer in \cite[\S5]{AxSi1} in the case where $A^c$ is acyclic; however
in our more general case new features arise and to deal with these we
derive new properties of the superfield propagator.
Throughout this section $A^c$ is an arbitrary flat gauge field which is 
isolated modulo gauge transformations (unless otherwise stated).

Axelrod and Singer used the BRS version of the usual Faddeev-Popov 
gauge-fixing procedure to derive a gauge-fixed expression for the 
Chern-Simons partition function as a functional integral over a superfield.
This expression can
be reproduced in our approach: We change variables in (\ref{4.17}) from
$\bar{C}\in\ker(d_0^{A^c})^{\perp}$ to $\bar{C}'=8{\pi}k^{-1}\lambda_{\bf g}
{\ast}d_0^{A^c}C\in\ker(d_2^{A^c})^{\perp}$ and define the superfield variable
\begin{eqnarray*}
\widehat{A}=C+\tilde{B}+\bar{C}'\in\ker(d_0^{A^c})^{\perp}\oplus\ker(d_1^{A^c})
^{\perp}\oplus\ker(d_2^{A^c})^{\perp}=\ker(d^{A^c})^{\perp}
\end{eqnarray*}
where $d^{A^c}$ denotes the covariant derivative on $\Omega(M,{\bf g})=
\oplus_{q=0}^3\Omega^q(M,{\bf g})\,$.
Substituting the resulting expression for (\ref{4.17}) in (\ref{4.19})
a straightforward calculation gives the following expression
for the gauge-fixed partition function:
\begin{eqnarray}
\lefteqn{I(\frac{1}{\sqrt{k}};1,-iS_{CS})_{{\lbrack}A^c\rbrack}} \nonumber \\
&=&V(H_{A^c})^{-1}\det(\tilde{\Delta}_0^{A^c})^{-1/2}
\det(8{\pi}k^{-1}\lambda_{\bf g}(\tilde{\Delta}_0^{A^c})^{-1/2})
e^{\frac{ik}{4\pi}S_{CS}(A^c)} \nonumber \\
& &\times\,\int_{\ker(d^{A^c})^{\perp}}{\cal D\/}\hat{A}e^{\frac{ik}
{4\pi}\int_M\mbox{Tr}(\hat{A}{\wedge}d^{A^c}\hat{A}+\frac{2}{3}
\hat{A}\wedge\hat{A}\wedge\hat{A})}
\label{7.1}
\end{eqnarray}
The functional integral in (\ref{7.1}), from which the higher order terms
in the perturbative expansion of the partition function are obtained, is the 
gauge-fixed expression obtained in 
\cite[(2.17)--(2.19)]{AxSi1}\footnote{No expression was given in 
\cite{AxSi1} for the overall factor multiplying the functional integral 
in (\ref{7.1}) or in the perturbative expansion (\ref{7.4}) below.}
with the requirement that $A^c$ is acyclic.
This requirement is the same as requiring $A^c$ to be isolated modulo gauge
transformations and weakly irreducible. The weak irreducibility means that
the isotropy subgroup $H_{A^c}$ is discrete, i.e. its Lie algebra
$\mbox{Lie}(H_{A^c})=\ker(d_0^{A^c})$ vanishes.
In our 
approach, using the refined version of the Faddeev-Popov procedure given
in \S4, the only requirement is that 
$A^c$ isolated modulo gauge transformations -in Chern-Simons gauge theory
this is equivalent to requiring $\mbox{Im}(d_0^{A^c})=\ker(d_1^{A^c})$.
Thus we see that our approach 
to gauge-fixing is equivalent to that of \cite{AxSi1} when their condition
is satisfied, and extends their approach to the case where $\ker(d_0^{A^c})$
is non-zero. For a number of simple 3-manifolds, e.g. $S^3$ and the lens
spaces, all the flat gauge fields $A^c$ have non-zero $\ker(d_0^{A^c})$ 
but satisfy our requirement $\mbox{Im}(d_0^{A^c})=\ker(d_1^{A^c})$.

The remainder of this section is devoted to showing that the terms in the
perturbative expansion of the Chern-Simons partition function are formally
metric-independent in our setting (i.e. with $A^c$ isolated modulo gauge
transformations). We will work with the expressions for these terms 
derived in \cite{AxSi1}. (The derivation of these expressions from (\ref{7.1})
goes through for arbitrary flat $A^c$). 
We begin by introducing the ingredients in these
expressions and the notations which are required to formulate them (see
\cite[\S3]{AxSi1} for more details). In the formulation given in \cite{AxSi1}
the propagator for the superfield $\hat{A}$ is taken to be a differential
form $L^{A^c}(x,y)$ on $M{\times}M$ with values in ${\bf g}\otimes{\bf g}^*$
defined as follows. The operator $d^{A^c}$ appearing in the quadratic term
in the exponential in (\ref{7.1}) restrict to an invertible map
$\tilde{d}^{A^c}:\ker(d^{A^c})^{\perp}\stackrel{\cong}{\to}\mbox{Im}(d^{A^c})$.
The propagator $L^{A^c}(x,y)$ is taken to be the
differential form version of the kernel-function for the operator
$\hat{L}^{A^c}:\Omega(M,{\bf g})\to\Omega(M,{\bf g})$ defined by
$\hat{L}^{A^c}=(\tilde{d}^{A^c})^{-1}$ 
on $\mbox{Im}(d^{A^c})$ and $\hat{L}^{A^c}=0$ on
$\mbox{Im}(d^{A^c})^{\perp}$. 
More precisely,
let $\{\rho^a\}_{a=1,\dots,\dim{\bf g}}$ be an orthonormal basis for 
${\bf g}$, then using the inner product in ${\bf g}$ to identify ${\bf g}^*$
with ${\bf g}$ we have
$L^{A^c}(x,y)=L_{ab}^{A^c}(x,y)\rho^a\otimes\rho^b\,$ given by 
\begin{eqnarray}
(\hat{L}^{A^c}\psi)^a(x)=\int_{M_y}L_{ab}^{A^c}(x,y)\wedge\psi^b(y)
\ \ \ \ ,\ \ \ \ \psi=\psi^b\rho^b\in\Omega(M,{\bf g})
\label{7.2}
\end{eqnarray}
where repeated indices are summed over. Here and in the following $M_x$ denotes
a copy of $M$ parameterised by a variable $x\,$, so for example we have
$L^{A^c}(x,y)\in\Omega^2(M_x{\times}M_y;{\bf g}\otimes{\bf g})$.
The propagator $L^{A^c}(x,y)$ diverges 
at the diagonal $x=y$ in $M_x{\times}M_y$
but is smooth away from the diagonal (see \cite[p.17--18]{AxSi1}).

We will be using the following general notations introduced in 
\cite[\S3]{AxSi1}: 
Each element $Q(x,y)=Q_{ab}(x,y)\rho^a\otimes\rho^b$ in $\Omega(M_x{\times}
M_y;{\bf g}\otimes{\bf g})$ corresponds to an element 
$Q_{ab}(x,y)\wedge\rho_{(x)}^a\wedge\rho_{(y)}^b$ in 
$\Gamma(M_x{\times}M_y;\Lambda((T^*M_x\oplus{\bf g}_x)\oplus(T^*M_y\oplus
{\bf g}_y)))$ where ${\bf g}_x$ and ${\bf g}_y$ are distinct copies of
${\bf g}\,$; this in turn determines an element 
\begin{eqnarray*}
Q_{tot}(x_1,\dots,x_V)\in\Gamma(M_{x_1}\times\cdots{\times}M_{x_V};
\Lambda(\oplus_{i=1}^V(T^*M_{x_i}\oplus{\bf g}_i)))
\end{eqnarray*}
defined by 
\begin{eqnarray}
Q_{tot}(x_1,\dots,x_V)=\sum_{i,j=1}^VQ_{ab}(x_i,x_j)\rho_{(i)}^a\rho_{(j)}^b
\label{7.3}
\end{eqnarray}
(Here and in the following we will often
omit the wedge symbol in wedge multiplication for notational convenience).
The perturbative expansion of (\ref{7.1}) derived in \cite{AxSi1} has 
the form
\begin{eqnarray}
I(\frac{1}{\sqrt{k}},1,-iS_{CS})_{{\lbrack}A^c\rbrack}=
Z_{sc}(k,A^c)\sum_{V=0,2,4,\dots}\Bigl(\frac{1}{\sqrt{k}}\Bigr)^VI_V(A^c)
\label{7.4}
\end{eqnarray}
(see \cite[(3.54)]{AxSi1}). The overall factor $Z_{sc}(k,A^c)$ multiplying
the series is the overall factor in (\ref{44.5})
with $D_{A^c}$ given by (\ref{3.11}) 
(this expression can also easily be obtained from (\ref{7.1})); the notation
reflects the fact that this factor is
the contribution from $A^c$ to the semiclassical approximation.
Using the techniques of \cite{AdSe(PLB)}, \cite[\S4.1]{AdSe(hep-th)} we find
\begin{eqnarray}
Z_{sc}(k,A^c)&=&
e^{-\frac{i\pi}{4}\eta({\ast}d_0^{A^c})}\biggl(\frac{4\pi\lambda_{\bf g}}{k}
\biggr)^{{\dim}H^0(d^{A^c})/2}
V_{\lambda_{\bf g}}(\tilde{H}_{A^c})^{-1}
V(M)^{-{\dim}H^0(A^c)/2} \nonumber \\
& &\times\,\tau(A^c)^{1/2}e^{\frac{ik}{4\pi}S_{CS}(A^c)}
\label{7.5}
\end{eqnarray}
where $\eta({\ast}d_0^{A^c})$ is the analytic continuation to 0 of the 
eta-function of ${\ast}d_0^{A^c}\,$,
$\tau(A^c)$ is the Ray-Singer torsion of $A^c$ \cite{RS}
and we have used the fact that $H_{A^c}$ can be identified with an invariant
subgroup $\tilde{H}_{A^c}$ of the gauge group $G\,$, from which it follows
that $V(H_{A^c})=V(M)^{{\dim}H^0(A^c)/2}V_{\lambda_{\bf g}}(\tilde{H}_{A^c})$
where $V_{\lambda_{\bf g}}(\tilde{H}_{A^c})$ is the volume of $\tilde{H}_{A^c}$
determined by the inner product in ${\bf g}$.
The product $V(M)^{-{\dim}H^0(A^c)}\tau(A^c)$ is metric-independent (a proof 
of this is given in \cite[\S4.1]{AdSe(hep-th)}) so the only metric dependence
of $Z_{sc}$ enters through the phase factor in (\ref{7.5}). ($Z_{sc}$ can be
made completely metric-independent by putting in by hand a phase factor
with phase given by Witten's geometric counterterm \cite[\S2]{W(Jones)}).

The coefficients $I_V(A^c)$ in the expression (\ref{7.4}) are given by
\cite[(3.54)]{AxSi1} to be
\begin{eqnarray}
I_V(A^c)=c_V\prod_{i=1}^V\Big\lbrack\int_{M_{x_i}}f_{a^ib^ic^i}
\frac{\partial}{\partial\rho_{(i)}^{a^i}}\frac{\partial}{\partial\rho_{(i)}
^{b^i}}\frac{\partial}{\partial\rho_{(i)}^{c^i}}\Big{\rbrack}L_{tot}^{A^c}
(x_1,\dots,x_V)^{\frac{3}{2}V}
\label{7.6}
\end{eqnarray}
where $c_V=(2{\pi}i)^{\frac{1}{2}V}((3!)^V(2!)^{\frac{3}{2}V}V!(\frac{3}{2}
V)!)^{-1}\,$, $\{f_{abc}\}$ are the structure constants of ${\bf g}$ given by
$\lbrack\rho^a,\rho^b\rbrack=f_{abc}\rho^c\,$, 
$\frac{\partial}{\partial\rho_{(i)}^a}$ is interior multiplication by 
$\rho_{(i)}^a$ and $L_{tot}^{A^c}(x_1,\dots,x_V)$ is defined as in (\ref{7.3}).
We choose the $\{\rho^a\}$ such that $f_{abc}$ is totally antisymmetric.
(The coefficient $I_V(A^c)$ can be interpreted as the contribution to
the perturbative expansion coming from all Feynman diagrams with $V$ vertices;
see \cite[p.22--24]{AxSi1} for the details).

The propagator $L_{ab}^{A^c}(x,y)$
can be expressed in terms of the eigenvectors and (non-zero)
eigenvalues for $D_{A^c}=\frac{1}{4\pi\lambda_{\bf g}}
{\ast}d_1^{A^c}$ and $\Delta_0^{A^c}$
in (\ref{44.1}) and (\ref{44.3}):
\begin{eqnarray}
L_{ab}^{A^c}(x,y)&=&-\frac{1}{4\pi\lambda_{\bf g}}\sum_j\frac{1}
{\lambda(j)}B_j^a(x){\wedge}B_j^b(y) \nonumber \\
& &+\sum_l\frac{1}{\mu(l)}\Bigl(\,C_l^a(x)\wedge({\ast}d_0^{A^c}C_l^b)(y)
-({\ast}d_0^{A^c}C_l^a)(x){\wedge}C_l^b(y)\Bigr)
\label{7.2.5}
\end{eqnarray}
where we have followed the convention of \cite[(3.53)]{AxSi1}.
Substituting the expression (\ref{7.2.5}) for the propagator into (\ref{7.6})
it is straightforward to verify that the perturbative expansion (\ref{7.4})
for the partition function coincides with the one obtained from our
``generalised momentum space'' formulation in \S5; this is a bit tedious
though so we omit the details.

As it stands the expression (\ref{7.6}) for $I_V$ is a formal expression.
Axelrod and Singer showed in \cite[\S3--\S4]{AxSi1} how it can be given
well-defined finite meaning, as we now discuss. 
The integrand in (\ref{7.6}) is not well-defined apriori: $L_{ab}^{A^c}(x,y)$
diverges on the diagonal $x=y$ so the terms $L_{ab}^{A^c}(x_i,x_i)\rho_{(i)}^a
\rho_{(i)}^b$ in $L_{tot}^{A^c}(x_1,\dots,x_V)$ are not well-defined.
However, as pointed out in \cite[p.17--18 and p.20]{AxSi1} the propagator
can be written as a sum of the form
\begin{eqnarray}
L_{ab}^{A^c}(x,y)=L^{A^c}(x,y)_{div}\delta_{ab}+L_{ab}^{A^c}(x,y)_{cont}
\label{7.7}
\end{eqnarray}
where $L^{A^c}(x,y)_{div}$ diverges on 
the diagonal $x=y$ and $L_{ab}^{A^c}(x,y)_{cont}$
is continuous across the diagonal. Since $\delta_{ab}\rho^a\wedge\rho^b=0$
we have $L_{ab}^{A^c}(x,y)\rho^a\rho^b
=L_{ab}^{A^c}(x,y)_{cont}\rho^a\rho^b$ for $x{\ne}y\,$,
which extends continuously across the diagonal $x=y$. Thus a well-defined 
expression for $L_{tot}^{A^c}(x_1,\dots,x_V)$ is obtained in a natural way by
replacing $L_{ab}^{A^c}(x_i,x_i)\rho_{(i)}^a\rho_{(i)}^b$ 
by $L_{ab}^{A^c}(x,y)_{cont}
\rho_{(i)}^a\rho_{(i)}^b$ for all $i=1,\dots,V$. In physics terminology
this can be interpreted as a point-splitting regularisation. With this
regularisation Axelrod and Singer showed that
each $I_V$ in the perturbative expansion (\ref{7.4}) is finite
\cite[theorem 4.2]{AxSi1}. (As pointed out in \cite[\S6, remark II(i)]{AxSi1}
the argument for this does not require any particular conditions on the flat
gauge field $A^c$). This remarkable result shows that with regard to 
perturbative expansion Chern-Simons gauge theory on compact 3-manifold is very
different from the usual quantum field theories in that no renormalisation
procedure is required to obtain finite expressions for the terms in the 
expansions. However, whether or not the perturbation series in (\ref{7.4})
converges is a completely different question which as far as we know has 
yet to be answered.

The definition of $L^{A^c}(x,y)$ requires a choice of metric $g$ on $M$ so the 
perturbative expansion (\ref{7.4}) is apriori metric-dependent. We noted in 
\S5 that the overall factor $Z_{sc}(k,A^c)$ given by (\ref{7.5})
is metric-independent (provided that a phase factor is put in by hand with
phase given by Witten's geometric counterterm). Thus any metric-dependence
of (\ref{7.4}) is contained in the coefficients $I_V(A^c)$ of the expansion.

We now establish the properties of the propagator $L^{A^c}(x,y)$ which
we will need to show the formal metric-independence of $I_V(A^c)$.
The first of these is
\begin{eqnarray}
L_{ba}^{A^c}(y,x)=-L_{ab}^{A^c}(x,y)
\label{7.8}
\end{eqnarray}
This is the property (PL3) in \cite[\S3]{AxSi1}; it can be derived for example
from the expression (\ref{7.2.5}) above. We define the spaces
${\cal H}_q^{A^c}\subset\Omega^q(M,{\bf g})$ by the orthogonal decompositions
$\ker(d_q^{A^c})=\mbox{Im}(d_q^{A^c})\oplus{\cal H}_q^{A^c}$ then the 
Hodge decomposition states 
\begin{eqnarray}
\Omega(M,{\bf g})=\mbox{Im}((d^{A^c})^*)\oplus\mbox{Im}(d^{A^c})\oplus
{\cal H}^{A^c}
\label{7.9}
\end{eqnarray}
where ${\cal H}^{A^c}=\oplus_{q=0}^3{\cal H}_q^{A^c}$. Let $\pi_
{d^{A^c}}\,$, $\pi_{(d^{A^c})^*}$ and $\pi_{{\cal H}^{A^c}}$
denote the orthogonal projections of $\Omega(M,{\bf g})$ onto
$\mbox{Im}(d^{A^c})\,$, $\mbox{Im}((d^{A^c})^*)$ and ${\cal H}^{A^c}$
respectively. Noting that $\mbox{Im}((d^{A^c})^*)=\ker(d^{A^c})^{\perp}$
it follows from the definitions that
\begin{eqnarray}
d^{A^c}\hat{L}^{A^c}=\pi_{d^{A^c}}\qquad,\qquad\hat{L}^{A^c}d^{A^c}=
\pi_{(d^{A^c})^*}
\label{7.10}
\end{eqnarray}
(as in \cite[(2.22)]{AxSi1}). Let $d_{M_x{\times}M_y}^{A^c}$ denote the 
covariant derivative on $\Omega(M_x{\times}M_y;{\bf g}\otimes{\bf g})$
determined by the flat gauge field $(A^c,A^c)$ on $M_x{\times}M_y\,$,
then a straightforward calculation using (\ref{7.9}) and (\ref{7.10})
gives
\begin{eqnarray}
d_{M_x{\times}M_y}^{A^c}L_{ab}^{A^c}(x,y)=(d_{M_x}^{A^c}+d_{M_y}^{A^c})
L_{ab}^{A^c}(x,y)=-(\delta_{ab}\delta(x,y)-\pi_{ab}^{A^c}(x,y))
\label{7.11} 
\end{eqnarray}
where $\delta(x,y)\in\Omega^3(M_x,M_y)$ is the differential form version of 
the kernel-function for the identity map on $\Omega(M)$ (as defined in 
\cite[(3.44)]{AxSi1}) and $\pi_{ab}^{A^c}(x,y)\in\Omega^3(M_x{\times}M_y)$
is the differential form version of the kernel-function for 
$\pi_{{\cal H}^{A^c}}$. (Here and in the following we are using the convention
defined in \cite[(3.53)]{AxSi1}). When $A^c$ is acyclic ${\cal H}^{A^c}=0$
and (\ref{7.11}) reduces to the property (PL1) stated in \cite[\S3]{AxSi1}.
We denote the variation of $L_{ab}^{A^c}(x,y)$ 
and $\pi_{ab}^{A^c}(x,y)$ under a variation ${\delta}g$ of the metric 
$g$ by $\delta_{{\delta}g}L_{ab}^{A^c}(x,y)$
and $\delta_{{\delta}g}\pi_{ab}^{A^c}(x,y)$. From (\ref{7.11}), using
the fact that $\delta_{ab}\delta(x,y)$ is metric-independent we obtain
\begin{eqnarray}
d_{M_x{\times}M_y}^{A^c}(\delta_{{\delta}g}L_{ab}^{A^c}(x,y))
=\delta_{{\delta}g}\pi_{ab}^{A^c}(x,y)
\label{7.12}
\end{eqnarray}
In the case which we are considering, i.e. where $A^c$ is isolated
modulo gauge transformations, we have ${\cal H}_0^{A^c}=\ker(d_0^{A^c})\,$,
${\cal H}_1^{A^c}={\cal H}_2^{A^c}=0$ and ${\cal H}_3^{A^c}=\ast{\cal H}_0
^{A^c}=\ast\ker(d_0^{A^c})$ so ${\cal H}^{A^c}=\ker(d_0^{A^c})\oplus
\ast\ker(d_0^{A^c})$. From now on we omit $A^c$ from the notation for the
sake of notational simplicity, setting $d_q=d_q^{A^c}\,$, $L_{ab}(x,y)=
L_{ab}^{A^c}(x,y)$ and $\pi_{ab}(x,y)=\pi_{ab}^{{\cal H}^{A^c}}(x,y)$.
The space $\ker(d_0)$ is independent of metric on $M$ and we can choose
a basis $\{h_i=h_i^a\rho^a\}_{i=1,\dots,\dim\ker(d_0)}$ for $\ker(d_0)\,$,
independent of metric, such that $<h_i(x),h_j(x)>_{\bf g}=\delta_{ij}$
for all $x{\in}M$. Using this basis we can write $\pi_{ab}(x,y)$ as
\begin{eqnarray}
\pi_{ab}(x,y)=h_i^a(x)h_i^b(y)V_g(M)^{-1}(vol_g(y)-vol_g(x))
\label{7.13}
\end{eqnarray}
where $V_g(M)$ and $vol_g$ are the volume and volume form of $M\,$, determined 
by the metric $g$ and orientation of $M$. In (\ref{7.13}) $vol_g(x)$ and
$vol_g(y)$ are the volume forms on $M_x$ and $M_y$ respectively, considered
as elements in $\Omega^3(M_x{\times}M_y)$.
There is a natural decomposition $\Omega(M_x{\times}M_y;{\bf g}\otimes
{\bf g})=\oplus_{p,q\in\{0,1,2,3\}}\Omega^{(p,q)}(M_x{\times}M_y;
{\bf g}\otimes{\bf g})$ where $\Omega^{(p,q)}(M_x{\times}M_y;{\bf g}\otimes
{\bf g})$ is the space of ${\bf g}\otimes{\bf g}$-valued forms of degree
p on $M_x$ and degree q on $M_y$. Using this we can write
\begin{eqnarray}
L_{ab}(x,y)&=&L_{ab}^{(0,2)}(x,y)+L_{ab}^{(1,1)}(x,y)+L_{ab}^{(2,0)}(x,y)
\label{7.14} \\
\pi_{ab}(x,y)&=&\pi_{ab}^{(0,3)}(x,y)+\pi_{ab}^{(3,0)}(x,y) \label{7.15} \\
\end{eqnarray}
where $L_{ab}^{(p,q)}(x,y)\in\Omega^{(p,q)}(M_x{\times}M_y)$ etc.
By substituting (\ref{7.14}) and (\ref{7.15}) into (\ref{7.12}) 
we see that
\begin{eqnarray}
d_{M_x{\times}M_y}(\delta_{{\delta}g}L_{ab}^{(1,1)}(x,y))&=&0 \label{7.16} \\
d_{M_x}(\delta_{{\delta}g}L_{ab}^{(0,2)}(x,y))=0\;\;\;\;&,&\;\;\;\;
d_{M_y}(\delta_{{\delta}g}L_{ab}^{(2,0)}(x,y))=0 \label{7.17}
\end{eqnarray}
Our condition on $A^c$ implies that the cohomology spaces $H^1(d)$ and
$H^2(d)$ vanish and it follows from the K\"unneth formula that 
$H^2(d_{M{\times}M})=0$. It then follows from (\ref{7.16}) and (\ref{7.8})
that
\begin{eqnarray}
\delta_{{\delta}g}L_{ab}^{(1,1)}(x,y)=d_{M_x{\times}M_y}B_{ab}(x,y)
\label{7.19}
\end{eqnarray}
for some $B(x,y)\in\Omega^1(M_x{\times}M_y;{\bf g}\otimes{\bf g})$ of the
form
\begin{eqnarray}
B_{ab}(x,y)=B_{ab}^{(0,1)}(x,y)-B_{ba}^{(0,1)}(y,x)
\label{7.20}
\end{eqnarray}
where $B_{ab}^{(0,1)}(x,y)\in\Omega^{(0,1)}(M_x{\times}M_y)$ satisfies
$d_{M_y}B_{ab}^{(0,1)}(x,y)=0$.
In our metric-independence argument below the properties 
(\ref{7.17})--(\ref{7.20}) of the propagator replace the key property
(PL4) in \cite[\S3]{AxSi1}.

The expression (\ref{7.6}) for $I_V(A^c)$ can be written compactly as in 
\cite[(3.55)]{AxSi1}:
\begin{eqnarray}
I_V(A^c)=c_V\int_{M^V}\mbox{TR}(L_{tot}(x_1,\dots,x_V)^{\frac{3}{2}V})
\label{7.21}
\end{eqnarray}
(recall from (\ref{7.4}) that $V$ is even) where $M^V=M_{x_1}\times\cdots
{\times}M_{x_V}$ and TR is a linear operator mapping $L_{tot}(x_1,\dots,x_V)
^{\frac{3}{2}V}$ to a differential form of top degree in $\Omega(M^V)$.
(TR is defined in \cite[p.21]{AxSi1}; it can be interpreted as a generalised
trace). Generalising the calculation in \cite[(5.83)]{AxSi1} we obtain
the following expression for the variation of $I_V(A^c)$ under a variation
${\delta}g$ of the metric:
\begin{eqnarray}
\lefteqn{\delta_{{\delta}g}I_V(A^c)} \nonumber \\
&=&\frac{3}{2}Vc_V\int_{M^V}\mbox{TR}((\delta_{{\delta}g}L_{tot})
(L_{tot})^{\frac{3}{2}V-1}) \nonumber \\
&=&\frac{3}{2}Vc_V\Big\lbrack\int_{M^V}\mbox{TR}((\delta_{{\delta}g}
L_{tot}^{(1,1)})(L_{tot})^{\frac{3}{2}V-1})+2\int_{M^V}\mbox{TR}((\delta_
{{\delta}g}L_{tot}^{(0,2)})(L_{tot})^{\frac{3}{2}V-1})\Big\rbrack 
\nonumber \\
&=&3Vc_V\Big\lbrack\int_{M^V}\mbox{TR}((d_{M^V}B_{tot}^{(0,1)})(L_{tot})^
{\frac{3}{2}V-1})+\int_{M^V}\mbox{TR}((\delta_{{\delta}g}L_{tot}^{(0,2)})
(L_{tot})^{\frac{3}{2}V-1})\Big\rbrack \nonumber \\
&=&3Vc_V\Big\lbrack\int_{M^V}\mbox{TR}((B_{tot}^{(0,1)})d_{M^V}(L_{tot})^
{\frac{3}{2}V-1})+\int_{M^V}\mbox{TR}((\delta_{{\delta}g}L_{tot}^{(0,2)})
(L_{tot})^{\frac{3}{2}V-1})\Big\rbrack \nonumber \\
&=&-\frac{3}{2}V(\frac{3}{2}V-1)c_V\Big\lbrack\int_{M^V}\mbox{TR}
(B_{tot}^{(0,1)}\delta_{tot}^{\bf g}(L_{tot})^{\frac{3}{2}V-2})
-\int_{M^V}\mbox{TR}(B_{tot}^{(0,1)}\pi_{tot}(L_{tot})^{\frac{2}{3}V-2})
\Big\rbrack \nonumber \\
& &+3Vc_V\int_{M^V}\mbox{TR}((\delta_{{\delta}g}L_{tot}^{(0,2)})
(L_{tot})^{\frac{3}{2}V-1})
\label{7.22}
\end{eqnarray}
where we have used 
\begin{eqnarray}
L_{tot}^{(0,2)}(x_1,\dots,x_V)=L_{tot}^{(2,0)}(x_1,\dots,x_V)
\label{7.23}
\end{eqnarray}
which follows from (\ref{7.3}) and (\ref{7.8}), and
\begin{eqnarray}
\delta_{{\delta}g}L_{tot}^{(1,1)}(x_1,\dots,x_V)=d_{M^V}B_{tot}(x_1,\dots,x_V)
=2d_{M^V}B_{tot}^{(0,1)}(x_1,\dots,x_V)
\label{7.24}
\end{eqnarray}
which follows from (\ref{7.3}), (\ref{7.19}) and (\ref{7.20}), and
\begin{eqnarray}
d_{M^V}L_{tot}(x_1,\dots,x_V)=-\delta_{tot}^{\bf g}(x_1,\dots,x_V)
+\pi_{tot}(x_1,\dots,x_V)
\label{7.25}
\end{eqnarray}
which follows from (\ref{7.3}) and (\ref{7.11}) with 
$\delta_{ab}^{\bf g}(x,y):=\delta_{ab}\delta(x,y)$.
In obtaining the fourth equality in (\ref{7.22}) we have used Stoke's theorem;
this requires $L_{tot}(x_1,\dots,x_V)$ to be a smooth form on $M^V$ which is 
not actually true since $L(x,y)$ diverges at $x=y$. Thus our calculation is
formal at this point\footnote{In \cite[\S5]{AxSi1}, \cite{AxSi2} Axelrod
and Singer gave a rigorous treatment of this problem (for $A^c$ acyclic).
They found that a metric-dependent phase factor appears, with phase given
by (minus) Witten's geometric counterterm. We are unsure as to whether
their argument for this continues to hold in our case.}; however 
at all other points (here and below) we are rigorous.
To show the formal metric-independence of $I_V(A^c)$ we show that the 3
integrals in (\ref{7.22}) all vanish. The first integral in (\ref{7.22})
has the form of the one appearing in the calculation of Axelrod and Singer
\cite[(5.83)]{AxSi1} and vanishes by the same argument which they gave.
(This involves cancellations between Feynman diagrams). The second and third
integrals in (\ref{7.22}),
\begin{eqnarray}
I_V^{(2)}&=&\int_{M^V}\mbox{TR}(\pi_{tot}B_{tot}(L_{tot})^{\frac{3}{2}V-2})
\label{7.26} \\
I_V^{(3)}&=&\int_{M^V}\mbox{TR}((\delta_{{\delta}g}L_{tot}^{(0,2)})(L_{tot})^
{\frac{3}{2}V-1}) \label{7.27}
\end{eqnarray}
did not arise in the calculation of Axelrod and Singer; they are new 
features of the more general situation which we are considering.
To show that $I_V^{(2)}$ and $I_V^{(3)}$ vanish we begin with some general
observations. The formula $d_q^*=-(-1)^{nq}{\ast}d_{n-q}\ast$ shows that
$\mbox{Im}(d^*)=\ast\mbox{Im}(d)$ and using this we get
\begin{eqnarray}
\int_M\phi^a\wedge\psi^a=0\qquad\mbox{for all}\;\;\phi=\phi^a\rho^a\in\ker(d^*)
\;,\;\;\psi=\psi^a\rho^a\in\mbox{Im}(d^*)
\label{7.28}
\end{eqnarray}
(with summation over repeated indices) by writing $\psi={\ast}d\tilde{\psi}$
and calculating
\begin{eqnarray*}
-\lambda_{\bf g}\int_M\mbox{Tr}(\phi\wedge\psi)=-\lambda_{\bf g}\int_M
\mbox{Tr}(\phi\wedge{\ast}d\tilde{\psi})=<\phi,d\tilde{\psi}>
=<d^*\phi,\tilde{\psi}>=0
\end{eqnarray*}
Recall from \S2 that the Lie bracket in ${\bf g}$ gives a Lie bracket in
$\Omega(M;{\bf g})$. The covariant derivative $d^{A^c}$ is a derivation
w.r.t. this bracket, and we have
\begin{eqnarray}
{\lbrack}h,\psi\rbrack\in\mbox{Im}(d^*)\qquad\mbox{for all}\;\;h\in\ker(d_0)
\,,\,\,\psi\in\mbox{Im}(d^*)
\label{7.29}
\end{eqnarray}
since, writing $\psi={\ast}d\tilde{\psi}\,$,
\begin{eqnarray*}
{\lbrack}h,\psi\rbrack={\lbrack}h,{\ast}d\tilde{\psi}\rbrack={\ast}{\lbrack}
h,d\tilde{\psi}\rbrack 
=\ast({\lbrack}dh,\tilde{\psi}\rbrack+{\lbrack}h,d\tilde{\psi}\rbrack)
={\ast}d{\lbrack}h,\tilde{\psi}\rbrack\in\mbox{Im}(d^*)
\end{eqnarray*}
Combining (\ref{7.29}) and (\ref{7.28}) gives
\begin{eqnarray}
\int_Mf_{abc}h^a\phi^b\wedge\psi^c=0\qquad\mbox{for all}\;\;h\in\ker(d_0)\;,
\;\phi\in\ker(d^*)\;,\;\psi\in\mbox{Im}(d^*)
\label{7.30}
\end{eqnarray}
(Note that (\ref{7.30}) holds in particular for $\phi\in\mbox{Im}(d^*)$ since
$\mbox{Im}(d^*)\subseteq\ker(d^*)\,$).
The operator $\hat{L}$ defining the propagator $L_{ab}(x,y)$ in (\ref{7.2})
has $\mbox{Im}(\hat{L})=\ker(d)^{\perp}=\mbox{Im}(d^*)\,$; it follows from this
and (\ref{7.8}) that 
\begin{eqnarray}
L_{ab}(x,y)\in\mbox{Im}(d_{M_x}^*)\qquad,{\qquad}L_{ab}(x,y)\in\mbox{Im}
(d_{M_y}^*)
\label{7.31}
\end{eqnarray}
(This is also easily derived from (\ref{7.2.5})).

We now use (\ref{7.30}) and (\ref{7.31}) to sketch how $I_V^{(3)}$ given 
by (\ref{7.27}) vanishes. Using (\ref{7.3}) $I_V^{(3)}$ can be expanded as
a sum of terms where each term involves an integral of the form
\begin{eqnarray}
\int_{M_y}f_{ace}(\delta_{{\delta}g}L_{ab}^{(0,2)}(y,x_i))L_{cd}(y,x_j))
L_{ef}(y,x_k)
\label{7.32}
\end{eqnarray}
(There are also terms where $L_{cd}(y,x_j)L_{ef}(y,x_k)$ is replaced by
$L_{ce}(y,y)$ in (\ref{7.32}) but these vanish since the integrand 
contains no 3-forms in $y$ in this case). From (\ref{7.17}) we have
$\delta_{{\delta}g}L_{ab}^{(0,2)}(y,x_i)\in\ker((d_{M_y})_0)\,$; 
combining this with (\ref{7.31}) we see from (\ref{7.30}) that (\ref{7.32})
vanishes\footnote{More precisely, the argument leading to (\ref{7.30})
generalises in an obvious way  to show that (\ref{7.32}) vanishes.}
so $I_V^{(3)}$ vanishes.

Finally, we sketch how $I_V^{(2)}$ given by (\ref{7.26}) vanishes.
Substituting the expression (\ref{7.13}) for $\pi_{ab}(x,y)$ into
$\pi_{tot}(x_1,\dots,x_V)$ in (\ref{7.26}) and expanding (\ref{7.26}) 
using (\ref{7.3}) leads to a sum of terms, each consisting of an integral
over $M^V$. A number of these terms vanish for one of the following 
reasons: \hfil\break
(i) $\pi_{ab}(x,x)=0$. (This follows from (\ref{7.13})). \hfil\break
(ii) The integrand in the integral over $M^V$ (a differential form on $M^V$)
is not of degree 3 in $x_i$ for all $i=1,\dots,V$. (Then the integral over 
$M_{x_i}$ vanishes). \hfil\break
(iii) The term contains an integral of the form
\begin{eqnarray}
\int_{M_y}f_{abd}h^a(y)L_{bc}(y,x_i)L_{de}(y,x_j)
\label{7.33}
\end{eqnarray}
which vanishes by (\ref{7.30}) since $h(y)\in\ker((d_{M_y})_0)$ and 
$L(y,x)\,,\;L(y,x_j)\in\mbox{Im}(d_{M_y}^*)$.
By inspection it is straightforward to check that the only terms which do not
vanish due to (i), (ii) or (iii) above are those of the form
\begin{eqnarray}
\lefteqn{\int_{M_z{\times}M_{x_i}{\times}M_{x_j}{\times}M_{x_k}{\times}M_y}
\Big\{
f_{acd}f_{bfp}h^a(y)B_{bc}^{(0,1)}(z,y)L_{de}^{(2,0)}(y,x_i)} \nonumber \\
& &\qquad\qquad\times\,L_{fg}(z,x_j)L_{pq}(z,x_k)\Psi_{egq}(x_i,x_j,x_k)\Big\}
\label{7.34}
\end{eqnarray}
or
\begin{eqnarray}
\lefteqn{\int_{M_z{\times}M_{x_i}{\times}M_{x_j}{\times}M_y}
\Big\{f_{ade}f_{bcg}
h^a(y)h^b(z)vol(z)B_{cd}^{(0,1)}(z,y)} \nonumber \\
& &\qquad\qquad\times\,L_{ef}^{(2,0)}(y,x_i)L_{gh}(z,x_j)\Phi_{fh}(x_i,x_j)
\Big\}
\label{7.34.5}
\end{eqnarray}
Therefore, to show that $I_V^{(2)}$ vanishes it suffices to show that
(\ref{7.34}) and (\ref{7.34.5}) vanishes. To show that (\ref{7.34}) vanishes
it suffices to show that
\begin{eqnarray}
\int_{M_y}f_{acd}h^a(y)B_{bc}^{(0,1)}(z,y)L_{de}^{(2,0)}(y,x_i)
\in\ker((d_{M_z})_0)
\label{7.35}
\end{eqnarray}
because then the integral over $M_z$ in (\ref{7.34}) vanishes by (\ref{7.30}).
To show (\ref{7.35}) we begin by noting that 
\begin{eqnarray}
\int_{M_y}f_{acd}h^a(y)L_{bc}^{(1,1)}(z,y)L_{de}^{(2,0)}(y,x_i)=0
\label{7.36}
\end{eqnarray}
for the same reason that (\ref{7.33}) vanished in (iii) above. Taking the 
metric-variation of this gives 
\begin{eqnarray}
0&=&\int_{M_y}f_{acd}h^a(y)(\delta_{{\delta}g}L_{bc}^{(1,1)}(z,y))L_{de}^
{(2,0)}(y,x_i)+\int_{M_y}f_{acd}h^a(y)L_{bc}^{(1,1)}(z,y)\delta_{{\delta}g}
L_{de}^{(2,0)}(y,x_i) \nonumber \\
&=&d_{M_z}\int_{M_y}f_{acd}h^a(y)B_{bc}^{(0,1)}(z,y)L_{de}^{(2,0)}(y,x_i)
+\int_{M_y}f_{acd}h^a(y)L_{bc}^{(1,1)}(z,y)\delta_{{\delta}g}L_{de}^{(2,0)}
(y,x_i) \nonumber \\
& &\label{7.37}
\end{eqnarray}
where we have used (\ref{7.19})--(\ref{7.20}). The first term in (\ref{7.37})
belongs to $\mbox{Im}(d_{M_z})$ while the second term belongs to
$\mbox{Im}(d_{M_z}^*)$ because of (\ref{7.31}). Since $\mbox{Im}(d^*)=
\ker(d)^{\perp}\subseteq\mbox{Im}(d)^{\perp}$ it follows that both terms in
(\ref{7.37}) vanish individually; the vanishing of the first term implies
(\ref{7.35}) so (\ref{7.34}) vanishes.
To show that (\ref{7.34.5}) vanishes we note that $h(z)vol(z)\in
\ker(d_{M_z}^*)$ since $d^*(h{\cdot}vol)={\ast}d{\ast}(h{\cdot}vol)=
{\ast}dh=0$. Combining this with (\ref{7.35}) and (\ref{7.31}) we see
that (\ref{7.34.5}) vanishes by (\ref{7.30}).
This completes the argument for the formal metric-independence of the 
coefficients $I_V(A^c)$ in the perturbative expansion of the partition
function.

\section{Conclusion}

We have described a method for carrying out a formal perturbative expansion
of the functional integral $I(\alpha;f,S)$ after 
expanding the action functional $S$ about a critical
point $A^c\,$, with the perturbative expansion being infrared-finite
when $A^c$ is isolated modulo gauge transformations. 
The main problem that we have solved in doing this is to carry out a 
gauge-fixing procedure of Faddeev-Popov type in such a way that infrared
divergences do not arise in the ghost propagator when 
$A^c$ is reducible\footnote{More
precisely, when $A^c$ is not weakly irreducible.}. This problem is particularly
relevant in Chern-Simons gauge theory on compact 3-manifolds, since for a
number of simple 3-manifolds such as $S^3$ and lens spaces all the flat
gauge fields $A^c$ are reducible. The usual Faddeev-Popov procedure (the
BRS version of which was used in \cite{AxSi1}) leads to the Faddeev-Popov
determinant
\begin{eqnarray}
\det\Bigl((d_0^{A^c})^*d_0^{A^c+B}\Bigr)
\label{c.1}
\end{eqnarray}
(with gauge fields $A=A^c+B\,$). Writing this as an integral over ghost 
fields $\bar{C}\,,\,C$ leads to the ghost term in the action functional
given by
\begin{eqnarray}
<\bar{C}\,,\,(d_0^{A^c})^*d_0^{A^c+B}C>\,=\,<\bar{C}\,,\,\Delta_0^{A^c}C>
+<\bar{C}\,,\,{\lbrack}B\,,\,C\rbrack>
\label{c.2}
\end{eqnarray}
Infrared divergences in the ghost propagator correspond to zero-modes in
the operator $\Delta_0^{A^c}=(d_0^{A^c})^*d_0^{A^c}$ in the quadratic
term in (\ref{c.2}); these are present when $A^c$ is reducible since
$\ker(\Delta_0^{A^c})=\ker(d_0^{A^c})=\mbox{Lie}(H_{A^c})$.
In our refinement of the Faddeev-Popov procedure, which takes account of
the ambiguities in the gauge-fixing condition coming from the
isotropy subgroup $H_{A^c}$ of $A^c\,$, we obtain
\begin{eqnarray}
V(H_{A^c})^{-1}\det\biggl(\,(d_0^{A^c})^*d_0^{A^c+B}\biggl|_{\ker(d_0^{A^c})^
{\perp}}\biggr)
\label{c.3}
\end{eqnarray}
instead of (\ref{c.1}). In this case the ghost propagator is 
infrared-finite for all $A^c$ since the operator $\Delta_0^{A^c}$ in 
(\ref{c.2}) is now restricted to the orthogonal complement of its
zero-modes. The appearance of the volume factor $V(H_{A^c})$ is also crucial
for a number of reasons. We saw in \S5 ((\ref{7.5}) and the subsequent
discussion) that this factor is necessary for metric-independence of the
of the overall factor multiplying the perturbation series for the
Chern-Simons partition function. It is also necessary for reproducing the
general formula \cite[app. II (9)]{Sch(Inst)} for the weak coupling limit
of $I(\alpha;f,S)$. Finally, the factor $V(H_{A^c})$ is necessary for
reproducing the numerical factors in the large $k$ limit of the expressions
for the partition function obtained from the non-perturbative prescription
of \cite{W(Jones)} from the semiclassical approximation. This was shown
in \cite{Roz} where the factors $V(H_{A^c})$ were put in by hand,
see also \cite[\S4.2]{AdSe(hep-th)}.

Our requirement that the spacetime manifold be compact riemannian without
boundary was important for avoiding infrared divergences in the perturbative 
expansions because it ensures that the operators $D_{A^c}$ and 
$\Delta_0^{A^c}$ have discrete spectra (at least for Yang-Mills-
and Chern-Simons gauge theories). If the spectra were continuous, with 
eigenvalues $\lambda(p)$ and $\mu(q)$ labelled by continuous parameters
$p$ and $q$ (as is the case e.g. in the usual flat spacetime setting where
$p$ and $q$ are momentum vectors) then the propagators, which are essentially
given by $\frac{1}{\lambda(p)}$ and $\frac{1}{\mu(q)}\,$, can be arbitrarily
large for sufficiently small $p$ and $q\,$, even though the values of
$p$ and $q$ for which $\lambda(p)=\mu(q)=0$ are excluded. This leads in
general to infrared divergences in the expressions for the Feynman diagrams.

A drawback with our method is that in order to explicitly evaluate the terms
in the perturbation series the eigenvectors and eigenvalues of $D_{A^c}$
and $\Delta_0^{A^c}$ must be determined. This is a non-trivial problem
in general (as opposed to the usual flat spacetime setting where it is
trivial).However, in Chern-Simons gauge theory with spacetime $S^3\,$, or $S^3$
divided out by the action of a finite group (e.g. a lens space), the 
eigenvectors and -values can be determined using the techniques of
\cite[\S4]{Ray}.

There are a number of interesting issues which are left unresolved in this 
paper. These are as follows:
\hfill\break
(1) We have seen that the gauge-fixing procedure can be carried out using
only the subgroup ${\cal G}_0$ of topologically trivial gauge transformations
(this is essentially because ${\cal G}_0$ and ${\cal G}$ have the same
Lie algebra), and that this avoids the usual problems that arise due to 
gauge-fixing ambiguities provided that our assumption in \S4 holds.
This assumption, that all ambiguities in the gauge-fixing come either
from $H_{A^c}$ or from topologically non-trivial gauge transformations,
should be verified (or disproved).
(2) The approach to perturbative expansion should be extended to obtain an
infrared-finite expansion in the general case where the critical point $A^c$ is
not isolated modulo gauge transformations. (We briefly discussed in \S4
how this might be done although our argument was incomplete. As mentioned
in the introduction S. Axelrod has recently announced a method for doing
this in Chern-Simons gauge theory when $A^c$ belongs to a smooth 
component of the modulispace of flat connections. This still leaves the
``non-generic'' case where ${\lbrack}A^c\rbrack$ is a singular point in
the modulispace; this case was discussed in the semiclassical approximation
in \cite{Roz}.)
\hfill\break
(3) The problem of ultraviolet divergences should be resolved 
(particularly for Yang-Mills theory) 
by extending the usual regularisation- and 
renormalisation procedures to the framework for perturbative expansion
given in this paper. The Pauli-Villars procedure and method of higher
covariant derivatives have a geometric nature which might make them suitable
for this.
\hfill\break
(4) The property of ``momentum conservation at the vertices'' 
of the Feynman diagrams in the usual flat spacetime setting 
should be generalised to our setting.
More precisely, the problem is to 
formulate and prove a theorem which describes how invariance
of the quadratic term $<B,D_{A^c}B>$ in the action functional under a
group of isometries of the spacetime manifold implies simplifying 
conditions analogous to momentum conservation at the vertices of the 
diagrams (cf. remark (ii) in \S2).
\hfill\break
(5) Perturbative expansion in Chern-Simons gauge theory is ultraviolet-finite
(after a point-splitting regularisation) due to a result in \cite{AxSi1}.
In our method the expansions are also infrared-finite when $A^c$ is isolated
modulo gauge transformations, so the terms in the perturbative expansion
of the Chern-Simons partition function are completely finite for a number
of simple 3-manifolds such as $S^3$ and lens spaces. It would be very 
interesting to explicitly calculate 
the terms in the expansions of the partition function for these
manifolds (using e.g. the techniques of \cite[\S4]{Ray}), determine whether
the perturbation series converges and see to what extent it reproduces the
expressions obtained from the non-perturbative prescription 
of \cite{W(Jones)}.

The expressions for the Chern-Simons partition function obtained from the
non-perturbative prescription of \cite{W(Jones)} with gauge group $SU(2)$
have been shown to agree with the semiclassical approximation 
in the limit of large $k$ for wide classes of
3-manifolds \cite{FG(PRL+CMP)} \cite{J} \cite{Roz} 
\cite{AdSe(PLB)} \cite{AdSe(hep-th)}. (However, no general proof of equality
between the semiclassical- and non-perturbative expressions in this limit
has been given so far\footnote{We thank M. Atiyah for emphasising this 
to us.}.) In showing this
the non-perturbative expressions were rewritten as a sum of terms
with each term corresponding to a flat gauge field $A^c$ (up to gauge 
equivalence), then in the large $k$ limit the term coincides with the 
lowest order term in the perturbative expansion determined by $A^c$.
This leads us to speculate that the full perturbation series determined
by $A^c$ may be equal to the non-perturbative term corresponding to $A^c$
for all values of the parameter $k\,$, in 
which case the complete non-perturbative 
expression is reproduced by evaluating the perturbative expansion determined
by $A^c$ for all the flat gauge fields $A^c$ on the 3-manifold and adding
these together. One detail to be dealt with before this could work out is the
fact that the non-perturbative expressions are analytic functions in
$\frac{1}{\sqrt{k+2}}$ (for gauge group $SU(2)\,$) 
rather than the coupling parameter $\frac{1}{\sqrt{k}}$.
However, this does not represent a serious problem because the non-perturbative
expressions can be rewritten as powerseries in $\frac{1}{\sqrt{k}}$.
In fact, for the cases we have looked at the resulting powerseries turns
out to have a surprisingly simple form \cite{unpublished}.

{\it Acknowledgements.} 
I am grateful to Siddhartha Sen for valuable discussions and encouragement
during the course of this work. I thank C.-M.~Viallet and O.~Babelon for
drawing my attention to \cite{Viallet1} \cite{Viallet(PLB)} \cite{Viallet2}.


\begin{thebibliography}{999}

\bibitem{Sch(Top)+NashSen}
A.~S.~Schwarz, Quantum field theory and topology
(Springer-Verlag, Berlin, 1993);
C.~Nash and S.~Sen, Topology and geometry for physicists (Academic Press,
London, 1983).

\bibitem{Sch(Baku)}
A.S.~Schwarz, Abstracts (Part II), Baku Int. Topological Conf. (Baku 1987).

\bibitem{W(Jones)}
E.~Witten, Commun. Math. Phys. 121 (1989) 351.

\bibitem{ItZu}
C.~Itzykson and J.-B.~Zuber, Quantum field theory 
(McGraw-Hill, New York, 1980).

\bibitem{Davies}
N.~Birrell and P.~Davies, Quantum fields in curved space
(Cambridge Univ. Press, Cambridge, 1982).

\bibitem{AxSi1}
S.~Axelrod and I.M.~Singer, Proc. XXth D.G.M. conf. (New York, 1991)
(S.~Catto and A.~Rocha, eds, World Scientific, 
Singapore and Teaneck NJ., 1992) pp. 3--45.

\bibitem{Guad}
E.~Guadagnini, The link invariants of the Chern-Simons field theory
(Walter de Gruyter, Berlin, 1993).

\bibitem{Natan}
D.~Bar-Natan, Princeton Univ. Ph.D. thesis, 1991.

\bibitem{AxSi2}
S.~Axelrod and I.M.~Singer, J. Differential Geom. 39 (1994) 173.

\bibitem{FP}
L.D.~Faddeev and V.N.~Popov, Phys. Lett. B 25 (1967) 29.

\bibitem{Rouet}
D.~Amati and A.~Rouet, Phys. Lett. B 73 (1978) 39.

\bibitem{Sch(Inst)}
A.S.~Schwarz, Commun. Math. Phys. 64 (1979) 233.

\bibitem{Sch(degen)}
A.S.~Schwarz, Commun. Math. Phys. 67 (1979) 1.   

\bibitem{Viallet1}
P.K.~Mitter and C.-M.~Viallet, Commun. Math. Phys. 79 (1981) 457.

\bibitem{talk}
Talk by the author at 1st Irish Q.F.T. Conf. (Dublin, 1994).

\bibitem{AdSe(hep-th)}
D.H.~Adams and S.~Sen, Trinity College Dublin
preprint TCD-95-03, hep-th/9503095.

\bibitem{Ax(hep-th)}
S.~Axelrod, M.I.T. preprint, hep-th/9511196.

\bibitem{Viallet(PLB)}
O.~Babelon and C.-M.~Viallet, Phys. Lett. B 85 (1979) 246.

\bibitem{Viallet2}
O.~Babelon and C.-M.~Viallet, Commun. Math. Phys. 81 (1981) 515.

\bibitem{Gilkey}
P.~Gilkey, Invariance theory, the heat equation, and the Atiyah-Singer
index theorem (Publish or Perish, Delaware, 1984).

\bibitem{unpublished}
Unpublished calculations by the author

\bibitem{BoossBleecker}
B.~Booss and D.~Bleecker, Topology and analysis: the Atiyah-Singer index
formula and gauge-theoretic physics (Springer-Verlag, New York, 1985).

\bibitem{Roz}
L.~Rozansky, Univ. of Texas at Austin preprints UTTG-06-93 hep-th/9303099
(to appear in Commun. Math. Phys.); UTTG-12-93 hep-th/9401060.

\bibitem{Sch(Top)}
A.S.~Schwarz, Topology for physists (Springer-Verlag, Berlin, 1993).

\bibitem{Fuchs}
J.~Fuchs, M.G.~Schmidt and C.~Schweigert, Nucl. Phys. B 426 (1994) 107.

\bibitem{Gribov}
V.~Gribov, Nucl. Phys. B 139 (1978) 1.

\bibitem{Jackiw(PRD)}
R.~Jackiw, I.~Muzinich and C.~Rebbi, Phys. Rev. D 17 (1978) 1576.

\bibitem{Singer}
I.M.~Singer, Commun. Math. Phys. 60 (1978) 7.

\bibitem{Nara}
M.S.~Narasimhan and T.R.~Ramadas, Commun. Math. Phys. 67 (1979) 21.

\bibitem{AdSe(PLB)}
D.H.~Adams and S.~Sen, Phys. Lett. B 353 (1995) 495.

\bibitem{Freed}
D.~Freed, Adv. Math. 113 (1995) 237.

\bibitem{Ray}
D.~Ray, Adv. Math. 4 (1970) 109.

\bibitem{RS}
D.~Ray and I.M.~Singer, Adv. Math. 7 (1971) 145.

\bibitem{FG(PRL+CMP)}
D.~Freed and R.~Gompf, Phys. Rev. Lett. 66 (1991) 1255; Commun. Math. Phys.
141 (1991) 79.

\bibitem{J}
L.~Jeffrey, Commun. Math. Phys. 147 (1992) 563


\end{thebibliography}
\end{document}